# Ytterbium charge state and stabilization in the Ba(Ca)$F_2$ host by electron paramagnetic resonance and infrared photoluminescence

David John[1,2*], Shelja Sharma[3*], Marius Stef[4], Gabriel Buse[4], Zdeněk Remeš[2], Anna Artemenko[2], Sergii Chertopalov[2], Vineet Sikarwar[3], Alan Mašláni[3], Jafar Fathi[3], Jakub Pilař[3], Tomáš Hostinský[3], Jan Zich[3], Tomáš Mates[3], Brenda Natalia Lopez Nino[3], Michal Hlína[3], Karol Bartosiewicz[2], Marina Konuhova[5], Anatoli Popov[5], Ján Lančok[2], Maksym Buryi[3*]

[1]*Faculty of Nuclear Sciences and Physical Engineering, Czech Technical University at Prague, Břehová 7, 115019, Prague, Czech Republic*
[2]*Institute of Physics of the Czech Academy of Sciences, Na Slovance 1999/2, 182 00, Prague, Czech Republic*
[3]*Institute of Plasma Physics of the Czech Academy of Sciences, U Slovanky 2525/1a, 182 00, Prague, Czech Republic*
[4]*Faculty of Physics, West University of Timisoara, Bd. Vasile Pârvan no. 4, 300223, Timisoara, Romania*
[5]*Institute of Solid State Physics, University of Latvia, Kengaraga street 8 LV-1063 Riga, Latvia*

**Abstract**

Lanthanide-doped fluorides are promising materials for advanced photonic and quantum applications due to their wide bandgap, low phonon energy, and chemical stability. In this work, we present a systematic comparative study of ytterbium incorporation at low doping levels (0.05-0.2 mol%) in $BaF_2$ and $CaF_2$ single crystals, focusing on the interplay between host lattice properties, charge-state stabilization, and defect formation mechanisms. Using a combination of X-ray diffraction (XRD), X-ray photoelectron spectroscopy (XPS), electron paramagnetic resonance (EPR), transmittance, and infrared photoluminescence (IR PL), we explore how host lattice properties affect the stabilization of $Yb^{3+}$ and $Yb^{2+}$ ions. XRD confirmed cubic phase purity and lattice parameter stability in both hosts, while XPS revealed surface chemical composition variations associated with charge-compensating defects and trace impurities. EPR spectra indicated that $BaF_2$ favored perturbed $Yb^{3+}$ environments with increasing dopant levels, while $CaF_2$ maintained predominantly unperturbed sites, suggesting a more favorable ionic match for $Yb^{2+}$. Photothermal deflection spectroscopy (PDS) and IR PL results showed host-specific optical responses, with $CaF_2$ exhibiting crystal-field splitting and broader local field effects. These results reveal a clear decoupling between long-range structural stability and local lattice perturbations, and demonstrate that host cation identity governs the balance between $Yb^{2+}$ and $Yb^{3+}$ stabilization as well as defect-driven optical behavior. This offers valuable insights for optimizing rare-earth-doped fluoride crystals in laser, scintillator, and quantum device applications.

*Keywords:* Ytterbium doping, charge state, rare-earth dopants, lattice interaction, X-ray photoelectron spectroscopy, electron paramagnetic resonance, lanthanide-doped fluorides.

*Corresponding authors: David John (john@fjfi.cvut.cz), Shelja Sharma (sharma@ipp.cas.cz), Maksym Buryi (e-mail: buryi@ipp.cas.cz)

## 1. Introduction

Lanthanide-doped fluorides have garnered significant interest owing to their remarkable optical and electronic properties, which render them highly suitable for applications in laser systems, radiation detectors, photovoltaics, and emerging quantum technologies [1,2]. Their intrinsic characteristics, such as wide bandgaps, low phonon energies, high UV transparency, and excellent chemical stability, facilitate the efficient incorporation of rare-earth ions, a critical factor in achieving high-performance luminescent and energy-converting materials [3,4]. Among the trivalent rare-earth ions, trivalent ytterbium ($Yb^{3+}$) holds a distinctive position. Its simple electronic configuration, $4f^{13}$, encompasses only two primary manifolds, $^2F_{7/2}$ and $^2F_{5/2}$, making it particularly suitable for applications in up-conversion, down-conversion, and near-infrared emission [5,6]. $Yb^{3+}$ ions demonstrate strong absorption around 980 nm and efficient emission in the range of 1000-1100 nm, with minimal multiphonon relaxation due to the substantial energy gap between their excited and ground states [7]. Additionally, the absence of closely spaced intermediate levels renders $Yb^{3+}$ a relatively straightforward model ion for investigating dopant-lattice interactions [8]. The behavior of Yb dopants within host matrices is intricate and is significantly influenced by local symmetry, compatibility of ionic radii, and mechanisms of charge compensation related to defects [9-11]. A key challenge in Yb-doped materials is the uncertainty surrounding its charge state. While $Yb^{3+}$ typically predominates, $Yb^{2+}$ can also emerge under specific conditions, particularly in materials where the host cation closely aligns with the ionic radius of $Yb^{2+}$ (1.14 Å). This can have substantial implications for the material's optical properties and stability [9,12].

In this context, alkaline-earth fluorides such as barium fluoride ($BaF_2$) and calcium fluoride ($CaF_2$) serve as particularly valuable host systems. Both crystallize in the cubic fluorite structure (space group Fm-3m) [10], offering a high-symmetry, eightfold coordinated environment that facilitates the analysis of dopant substitution and lattice distortion. However, their cationic sizes and polarizabilities differ significantly: $Ba^{2+}$ has a larger ionic radius (approximately 1.42 Å), while $Ca^{2+}$ is notably smaller (about 1.12 Å) [11]. This variation not only influences the lattice strain caused by Yb substitution but also affects the likelihood of various defect compensation pathways and the stabilization of specific Yb valence states. Particularly, substituting $Ba^{2+}$ or $Ca^{2+}$ with $Yb^{3+}$ (0.985 Å) introduces both a net positive charge and a size discrepancy, which necessitates local structural or electronic compensation [13]. Charge neutrality can be restored through the creation of cation vacancies ($V_m$), interstitial fluorine ($F_i$), or complex defect clusters. In $BaF_2$, the incorporation of $Yb^{3+}$ is frequently associated with interstitial $F^-$ defects and cationic vacancies [14]. In contrast, the closer size match between $Yb^{2+}$ and $Ca^{2+}$ may promote a higher concentration of divalent ytterbium ions, thereby reducing the need for compensating defects. First-principles calculations by Kye et al. confirmed that fluorine-rich environments favor the formation of $F_i$ defects in $CaF_2$, effectively stabilizing the charge-compensated $Yb^{3+}$ sites while suppressing non-radiative centers [8]. Experimental results utilizing photoluminescence (PL), X-ray absorption spectroscopy (XAS/EXAFS), and EPR indicate that the equilibrium between $Yb^{3+}$ and $Yb^{2+}$ in fluoride hosts such as $CaF_2$ and $BaF_2$ is significantly influenced by factors beyond the host lattice itself. Research has demonstrated that dopant concentration plays a crucial role in affecting dopant clustering and luminescent centers. At low concentrations, $Yb^{3+}$ remains isolated within high-symmetry cubic sites; however, as the concentration increases, cluster centers (including dimers, trimers, and larger configurations) begin to form, thereby altering charge compensation and emission characteristics. These findings have been corroborated by PL and EPR analyses of up-converting $CaF_2$:$Yb^{3+}$/$Er^{3+}$ nanoparticles [11]. Moreover, ambient conditions and impurity presence play a critical role. XAS/EXAFS studies (e.g. Er in $CaF_2$) found that higher dopant loading promotes clustering associated with Ca vacancies, influencing the local oxidation and coordination environment [15]. In particular, environmental exposure to oxygen and halogens like $Cl^-$ can result in surface-bound complexes or defect states in Yb-doped fluoride materials. This was demonstrated by Nelson et al. in their study of Yb-doped strontium fluoroapatite, where high-resolution XPS revealed distinct O 1s features corresponding to adsorbed oxygen that were separate from lattice oxygen. Additionally, shifts in the F 1s binding energy indicated modified halide coordination environments. These alterations in the local electronic structure surrounding the Yb dopants significantly affect charge state and coordination [16].

In fluoride-based hosts such as $CaF_2$ and $BaF_2$, the dominant intrinsic defects are related to the anion sublattice, in particular fluorine vacancies and associated defect complexes [17-19]. These defects introduce localized electronic states within the bandgap, which can influence the luminescence properties of $Yb^{3+}$ ions. In general, fluorine vacancies act as trapping centers that may facilitate non-radiative relaxation processes, leading to partial quenching of the NIR emission [17,18]. At the same time, they can locally modify the crystal field surrounding $Yb^{3+}$ ions, resulting in slight variations in emission intensity, linewidth, and spectral position [19].

Besides fluorine vacancies, several other types of defects can be expected in these materials. These include interstitial fluorine ions, cation vacancies ($Ca^{2+}$ or $Ba^{2+}$ vacancies), and defect complexes formed for charge compensation, particularly in the presence of trivalent dopants such as $Yb^{3+}$ [18,20]. In such cases, local charge imbalance may lead to the formation of $Yb^{3+}$-vacancy associates or more complex defect clusters. Furthermore, impurity-related defects (e.g., oxygen contamination or hydroxyl groups) and radiation-induced color centers may also be present, especially under non-ideal growth conditions or irradiation [17,19,21]. These defects can act either as non-radiative recombination centers or as perturbations of the local crystal field, thereby affecting the optical response.

Therefore, although defect-related processes cannot be completely neglected, their influence in $CaF_2$ and $BaF_2$ single crystals remains moderate and does not compromise the intrinsic advantages of these materials, such as low phonon energy, well-defined lattice environment, and efficient $Yb^{3+}$ NIR emission.

A wide range of strategies has been explored to enhance near-infrared (NIR) emission in $Yb^{3+}$-activated materials. Among them, perovskite nanocrystals represent the most efficient systems, where quantum cutting enables apparent photoluminescence quantum yields (PLQY) exceeding 100%. For instance, $CsPbCl_3$:$Yb^{3+}$ and $CsPb(Cl_{1-x}Br_x)_3$:$Yb^{3+}$ nanocrystals exhibit PLQY values up to ~200% due to the generation of two NIR photons per absorbed high-energy photon [22]. Similarly, $CsPbCl_3$:$Yb^{3+}$ thin films show PLQY values above 60% and external quantum efficiency (EQE) up to ~5.9% in device configurations [23]. Surface engineering further improves the emission properties, as demonstrated for $CsPbBr_3$:$Yb^{3+}$ nanocrystals, where PLQY increases from ~44% to ~64% after $SiO_2$ encapsulation [24]. Encapsulation of nanocrystals also leads to the improved environmental and luminescent properties [25].

Another commonly used approach involves sensitization via energy transfer processes. In systems such as $Na_5Y(WO_4)_4$:$Nd^{3+}$,$Yb^{3+}$ - $Yb^{3+}$ emission is not observed under direct excitation and appears only via $Nd^{3+} \rightarrow Yb^{3+}$ energy transfer, with efficiencies in the range of 50-75% [26]. However, such multi-component systems rely on indirect excitation pathways and therefore do not provide intrinsic emission efficiency of $Yb^{3+}$ itself.

In contrast, molecular systems typically exhibit significantly lower efficiencies. $Yb^{3+}$-based organic complexes show quantum yields in the range of only 0.5-3.8%, mainly due to strong non-radiative relaxation induced by high-energy vibrational modes of the ligands [27]. This highlights the importance of the host matrix in controlling radiative versus non-radiative processes.

For conventional inorganic hosts, including fluorides, nitrides, oxides, and glass-based systems, the situation is markedly different. Materials such as $CaF_2$:$Yb^{3+}$ [28], $La_3Si_6N_{11}$:$Yb^{3+}$ [29], $MgGeO_3$:$Yb^{3+}$ [5], $Y_3Ga_5O_{12}$:Yb [30] (in mixed garnets Ga content influences rare-earth ions incorporation and luminescence [31]), and $Yb^{3+}$-doped glass or glass-ceramics [32] consistently exhibit characteristic NIR emission in the ~970-1030 nm range. However, in most of these systems, absolute photoluminescence quantum yield or output power is not reported. Instead, performance is typically described using alternative parameters such as emission cross-sections, lifetimes, persistent luminescence, or laser threshold values. This lack of directly comparable efficiency metrics represents a general limitation in the literature.

Luminescent glasses differ fundamentally from crystalline systems due to their disordered structure. This effect is well documented in rare-earth systems, where significantly lower quantum yields are observed in glasses compared to crystalline hosts due to stronger non-radiative processes and less rigid coordination environments [33]. In addition, impurities such as $OH^-$ groups act as efficient quenching centers, enabling non-radiative de-excitation via high-energy vibrational modes and further reducing luminescence efficiency [34]. In glass hosts, $Yb^{3+}$ emission bands are significantly broadened, and the local environment varies continuously, leading to less well-defined optical transitions [32]. Moreover, glasses typically possess higher phonon energies compared to fluoride crystals, which enhances multiphonon relaxation and reduces emission efficiency. For example, in silica glass, the spectroscopic performance is constrained by low rare-earth solubility, which leads to clustering of $Yb^{3+}$ ions and subsequent concentration quenching, resulting in reduced fluorescence lifetime and non-exponential decay behavior at higher doping levels [35]. As a result, their performance is often evaluated using spectroscopic parameters rather than quantum yield.

Glass-ceramic systems partially overcome these drawbacks by introducing nanocrystalline phases into the amorphous matrix. These crystalline inclusions provide more defined local environments for $Yb^{3+}$ ions, leading to increased emission intensity, broader emission bandwidth, and longer lifetimes compared to the parent glass [36]. Nevertheless, even in such hybrid systems, the spectroscopic performance remains inferior to that of fully crystalline materials.

In this context, $CaF_2$ and $BaF_2$ single crystals represent a particularly advantageous class of materials. Their low phonon energy suppresses non-radiative losses, while the well-defined crystal field provides stable and reproducible emission centered around ~980 nm. Unlike perovskite or sensitized systems, $Yb^{3+}$ emission in fluorides is intrinsic and does not rely on complex multi-step mechanisms. In addition, these materials exhibit high chemical, thermal, and radiation stability, making them suitable for demanding environments. Although they do not reach the extremely high apparent efficiencies of quantum cutting systems, $CaF_2$ and $BaF_2$ offer a unique combination of simplicity, stability, and physical reliability, which is highly desirable for practical applications. Despite extensive studies on rare-earth-doped fluorides, a direct and systematic comparison between $BaF_2$ and $CaF_2$ hosts at low Yb concentrations, combining structural, spectroscopic, and defect-sensitive techniques, remains largely unexplored. In particular, the interplay between local lattice distortion, charge-state stabilization, and surface defect chemistry has not been comprehensively addressed.

In this work, we present a comparative investigation of Yb incorporation behavior in $BaF_2$ and $CaF_2$ single crystals across low doping concentrations (0.05-0.2 mol%). By employing a combination of X-ray diffraction, electron paramagnetic resonance, X-ray photoelectron spectroscopy, photoluminescence, and Raman spectroscopy, we assess the influence of host cation identity on Yb charge state, local environment, and defect evolution. Our goal is to elucidate how host lattice properties affect $Yb^{3+}$ and $Yb^{2+}$ stabilization, and how these, in turn, impact optical and electronic properties relevant to photonic and quantum applications.

## 2. Experimental

### 2.1 Samples preparation

$YbF_3$-doped $BaF_2$ and $CaF_2$ single crystals were grown using the vertical Bridgman technique, a well-established method for the growth of high-quality rare-earth-doped fluoride single crystals [37,38]. In both cases, high-purity alkaline-earth fluorides ($BaF_2$ or $CaF_2$, suprapure grade, Merck) and ytterbium fluoride ($YbF_3$, 99.99%, Merck) were used as starting materials. The $YbF_3$ dopant concentration in the melt was varied in the range of 0.05-0.2 mol%. The powders were weighed according to the nominal composition, thoroughly mixed to ensure uniform dopant distribution, and loaded into graphite crucibles (approximately 10 cm in length and 10 mm in diameter). The resulting crystals exhibited good optical transparency and mechanical integrity, with no visible cracks or secondary phases. The entire growth process was performed under high-vacuum conditions (typically better than $10^{-1}$ Pa), which suppress hydrolysis reactions and minimize oxygen- and moisture-related defects in the melt.

For both $BaF_2$:Yb and $CaF_2$:Yb crystals, the growth procedure involved four steps: (i) heating to complete melting of the charge; (ii) a melt homogenization stage lasting 3 hours; (iii) directional solidification at a constant pulling rate of about 4 mm/h; (iv) after completion of the growth process, the crystals were cooled to room temperature under controlled conditions, with a typical cooling rate of approximately 3 °C/min, in order to minimize thermal stress and dislocation formation. The crystals were cleaved along the (111) crystallographic planes into optically transparent slices with thicknesses ranging from approximately 1.5 to 3 mm, suitable for subsequent structural, optical, and spectroscopic investigations.

The use of identical growth parameters (crucible geometry, vacuum conditions, pulling rate, and thermal profile) ensures that the observed differences arise primarily from intrinsic host lattice effects rather than synthesis variations. Crystal growth was performed using the vertical Bridgman technique under comparable thermal conditions for both systems. The melting temperatures were approximately 1368 °C for $BaF_2$ and 1418 °C for $CaF_2$. The total growth duration was in the range of 36 hours, including the melting, homogenization, directional solidification, and cooling stages. The solidification process was carried out at a pulling rate of ~4 mm/h, followed by controlled cooling to room temperature to minimize thermal stress. The samples investigated were obtained from the cleaved as-grown crystals as shown in Fig. 1.

All samples were grinded into powders under inert conditions in a nitrogen-filled glovebox using a ball mill to prevent oxidation and moisture contamination prior to characterization. The resulting powders were subsequently sieved to obtain particles with sizes below 50 μm. This approach was primarily motivated by the requirements of electron paramagnetic resonance (EPR) measurements, where powder samples enable orientation averaging and reliable determination of spin-Hamiltonian parameters. In addition, for irradiation experiments, the increased surface-to-volume ratio of powders enhances the formation and sensitive detection of radiation-induced defect centers.

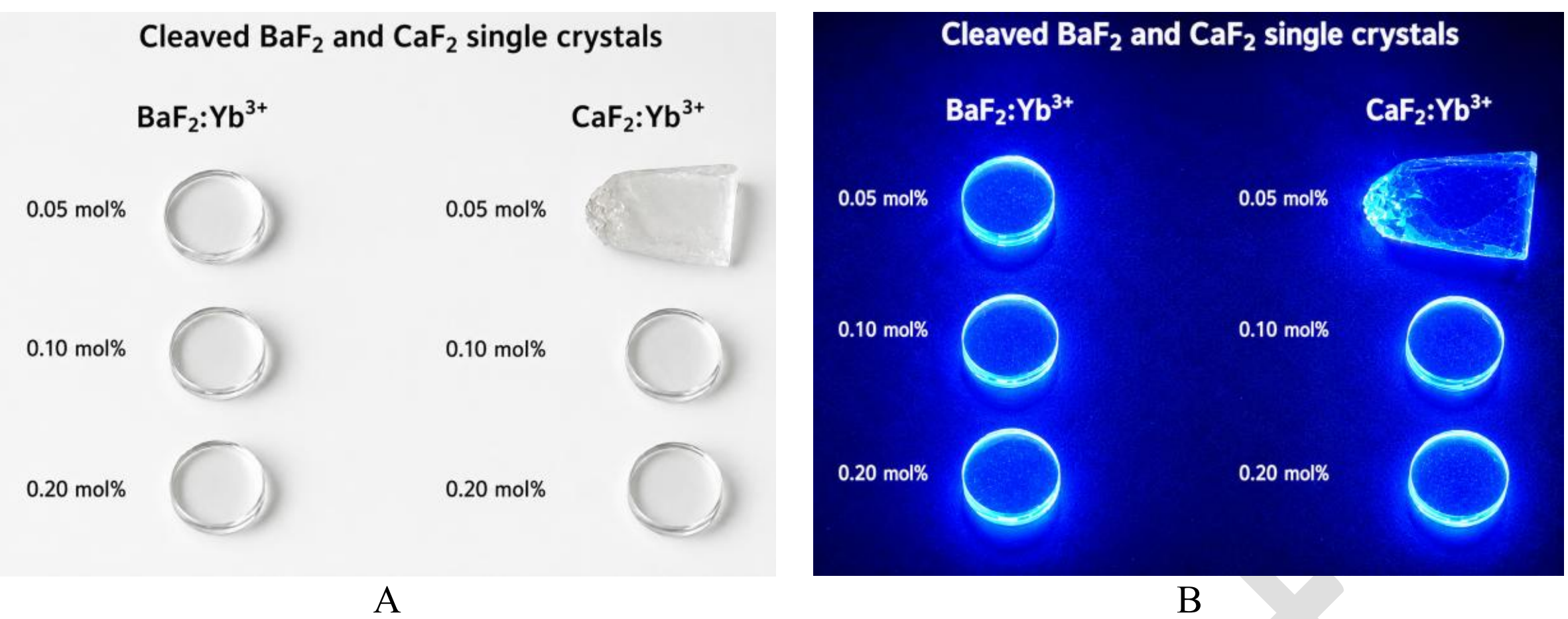

A B

Fig. 1. Photographs of the cleaved $BaF_2$:$YbF_3$ and $CaF_2$:$YbF_3$ single crystals grown by the vertical Bridgman method for different $YbF_3$ concentrations (0.05, 0.10, and 0.20 mol%) under ambient illumination (A) and UV excitation (B).

The use of powdered samples is also standard for X-ray diffraction (XRD) measurements, providing statistically representative structural information. To ensure consistency across different experimental techniques all measurements were therefore performed on powdered samples.

2.2 Experimental techniques

X-ray diffraction patterns were recorded in the $2\theta$ range of 10-120° using an Empyrean (Malvern, Panalytical) diffractometer with Cu $K_{<\alpha>}$ radiation (λ = 0.15418 nm), operating at $U$ = 45 kV, $I$ = 30 mA, with a step size of 0.013° and a counting time of 100 s per step. Phase identification was carried out using HighScore Plus 4.6a XRD analysis software. Powders used for XRD measurements were obtained by crushing the grown single crystals. The characterization of powders was performed in the Bragg-Brentano geometry using a fixed divergence slit with an anti-scatter slit, a 1D scanning line detector, and a Ni filter. To avoid reflections from the sample holder, powders were placed on a Si single-crystal wafer with a definite cut. The ICDD PDF-2 database, version 2013, was used for comparison. To complement the XRD data, scanning electron microscopy (SEM) was used to assess sample shape and size on a MAIA3, TESCAN electron microscope. The in-beam scanning electron (SE) detector is situated in the objective lens. The energy of the electron beam was 5 keV.

The transmission spectra were measured in the 300-1400 nm spectral range using Xe lamp as a light source, SpectraPro-150 monochromator (gratings: UV holographic 1200/mm and a ruled 600/mm blazed at 500 nm) and integrating sphere equipped with Si and InGaAs photodiodes as detectors. The spectral resolution was 5 nm with the UV grating and 10 nm with the ruled grating [39]. To suppress the light scattering, the samples were immersed into UV transparent liquid (Fluorinert FC72)

Infrared transmission and Raman spectra were recorded using a Fourier-transform Infared (FTIR) spectrometer (Thermo Scientific Nicolet™ iS50 FTIR Spectrometer) equipped with Nd:YAG laser (500 mW, 1064 nm) and FTIR Raman module. The scattered radiation was collected in a backscattering configuration and analyzed using an interferometer-based system. The signal was detected using an DTGS (transmission) and InGaAs (Raman) detector. The use of 1064 nm excitation effectively suppresses fluorescence background, which is particularly important for rare-earth-doped fluoride materials. The spectral resolution was 4 $cm^{-1}$ (transmission) and 2 $cm^{-1}$ (Raman). All measurements were performed at room temperature under ambient conditions.

EPR measurements were done on a commercial Bruker EMXplus spectrometer (X-band with the 9.4 GHz frequency) having sensitivity of around $10^{12}$ spins/mT. The temperature range was 10-60 K (Oxford Instruments ESR900 cryostat was used). The experimental spectra were fitted with the calculated ones using Easyspin 6.0.12 comprehensive software [40].

X-ray photoelectron spectroscopy was used for the investigation of surface chemical composition. The XPS spectra were acquired with AXIS Supra photoelectron spectrometer (Kratos Analytical Ltd., UK) equipped with a monochromated Al $K_\alpha$ X-ray source (1486.6 eV) and hemispherical energy analyzer. The XPS survey spectra were collected at a pass energy of 80 eV, whereas the high-resolution spectra were acquired at the pass energy of 20 eV. XPS spectra were obtained at a constant take-off angle of 90˚ from the analysis area of $0.3 \times 0.7$ mm$^2$. The charge neutralization system was used during XPS measurements. Peak fitting of the measured high resolution (HR) spectra was performed by the CasaXPS software using Shirley background and Gaussian/Lorentzian line shapes (GL(30), 70% Gaussian/30% Lorentzian) without fixing of peak full width at half maximum (FWHM). Binding energies (BE) of photoelectron lines were determined with the accuracy ± 0.2 eV. The obtained XPS spectra for $BaF_2$ and $CaF_2$ samples were calibrated on Ba 3d at 780.6 eV and Ca 2p at 346.7 eV, respectively [43]. For XPS analysis, the samples were fixed with a Cu tape on a sample holder. Elemental concentrations were determined by the integration of high resolution XPS peak spectra using Linear background and relative sensitivity factors. No procedure was used for samples surface cleaning inside XPS spectrometer prior to the XPS measurement.

## 3. Results and discussion

The comparison presented in Table 1 clearly demonstrates that although perovskite nanocrystals exhibit exceptionally high photoluminescence quantum yields (up to ~200%), they suffer from strong power saturation and environmental instability. In contrast, $Yb^{3+}$-doped fluoride single crystals such as $CaF_2$ and $BaF_2$ exhibit stable emission around 980 nm with long excited-state lifetimes (~ms range) and negligible non-radiative losses. Unlike nanocrystalline or molecular systems, these bulk fluoride hosts are free of surface-related quenching and degradation effects. Furthermore, compared to oxide hosts such as YAG, fluorides provide lower phonon energies, which is beneficial for efficient NIR emission. Therefore, $BaF_2$ and $CaF_2$ single crystals represent a highly robust and efficient platform for NIR emission, particularly suitable for radiation detection and high-energy applications.

Table 1. Comparison of $Yb^{3+}$-based materials: NIR emission peaks and reported performance parameters.

| Material | NIR emission peak | Power / efficiency | Ref. |
|---|---|---|---|
| $Y_3Al_5O_{12}$:$Yb^{3+}$ (YAG:Yb) | ~1053 nm (laser emission); absorption ~940 nm | Slope efficiency up to ~32% (10 at.% $Yb^{3+}$), decreasing to ~26% at higher doping | [42] |
| $Er^{3+}$/$Yb^{3+}$(/$Bi^{3+}$):$Cs_2NaYCl_6$ (double perovskite) | ~980 nm ($Yb^{3+}$ sensitizer); ~1530 nm ($Er^{3+}$ emission) | Luminescence enhancement ~2.5×; temperature sensitivity (not PLQY (Photoluminescence Quantum Yield)) | [43] |
| $[Yb(az)_4]^-$ (organic complex) | ~980 nm | Quantum yield ~0.5-1 % | [27] |
| $[NaYb(2a)_4]$ (organic complex) | ~980 nm | Quantum yield up to ~3.8 % | [27] |
| $[Yb(hfth)_3phen]$ (organic complex) | ~980 nm | Quantum yield ~1-2 % | [27] |
| [Yb(5a)(H5a)] (organic complex) | ~980 nm | Quantum yield ~0.5-1 % | [27] |
| $CsPbCl_3$:$Yb^{3+}$ nanocrystals | ~990 nm | PLQY up to ~200% (quantum cutting) | [22] |
| $CsPb(Cl_{1-x}Br_x)_3$:$Yb^{3+}$ nanocrystals | ~980-990 nm | PLQY ≈ 200% (quantum cutting) | [22] |
| $CsPbCl_3$:$Yb^{3+}$ thin film | 984 nm | PLQY >60%; EQE ≈ 5.9% | [23] |
| $CsPbBr_3$:$Yb^{3+}$ nanocrystals | 985 nm | NIR PLQY ≈ 44% | [24] |
| $CsPbBr_3$:$Yb^{3+}$@$SiO_2$ nanocrystals | 985 nm | NIR PLQY up to ≈ 64% | [24] |
| $CaF_2$:$Yb^{3+}$ | ~980 nm | Not reported | [28] |
| $Yb^{3+}$-doped cyclen complexes | ~980 nm | Not reported | [32] |
| $La_3Si_6N_{11}$:$Yb^{3+}$ | 983 nm | Not reported | [29] |

| $MgGeO_3:Yb^{3+}$ | ~980 nm, 1019 nm | Not reported (persistent luminescence reported) | [5] |
|---|---|---|---|
| $Na_5Y(WO_4)_4:Nd^{3+},Yb^{3+}$ | ~1011 nm | Energy transfer efficiency 50-75% (Nd→Yb), no absolute PLQY | [26] |
| $Y_3Ga_5O_{12}:Yb^{3+}$ | 971, 1001, 1025 nm | Laser threshold Imin ≈ 2.73 $kW\cdot cm^{-2}$ (no PLQY) | [30] |
| $Cs_2NaYbCl_6$ | ~980 nm | PLQY ≈ 43.6% | [44] |

Unlike many previously reported systems, $BaF_2$:Yb and $CaF_2$:Yb combine intrinsic $Yb^{3+}$ emission, structural simplicity, radiation stability, and low-phonon-energy fluoride hosts without relying on sensitization or quantum-cutting mechanisms.

It should also be noted that for many conventional $Yb^{3+}$-activated inorganic hosts, absolute photoluminescence quantum yield or output power is not reported. Instead, performance is typically described using alternative parameters such as energy-transfer efficiency, emission cross-sections, lifetimes, or laser threshold values.

Based on the comparison of the literature, it is evident that although $Yb^{3+}$ exhibits consistent NIR emission around ~980 nm across a wide range of materials, the reported performance metrics vary significantly. Perovskite systems can achieve very high PLQY values (up to ~200%) via quantum cutting, but they suffer from limited chemical and thermal stability. Organic complexes show low quantum yields (<4%) and poor robustness, while in most inorganic hosts (oxides, nitrides, glasses, ceramics) quantum yield or output power is often not reported at all, with performance instead described by indirect parameters such as lifetimes, emission cross-sections, or laser thresholds.

In this context, $CaF_2$ and $BaF_2$ single crystals represent a particularly advantageous platform, as they combine the simple and well-defined electronic structure of $Yb^{3+}$ with a low-phonon host that suppresses non-radiative losses. Unlike complex or sensitized systems, the emission is intrinsic and does not rely on multi-step mechanisms, leading to reproducible and physically well-understood behavior. At the same time, these fluoride matrices exhibit high chemical, thermal, and radiation stability. Although they do not reach the extreme PLQY values of some perovskites, they offer a unique combination of stability, simplicity, and reliability, which is essential for practical applications.

*3.1 Phase purity*

XRD analysis demonstrates that the $BaF_2$:Yb and $CaF_2$:Yb samples are single-phase materials with cubic $BaF_2$ and $CaF_2$ crystal structures (space group *Fm-3m*, reference codes 96-720-3815 and 00-035-0816, respectively). The corresponding XRD patterns are shown in Fig. 2.

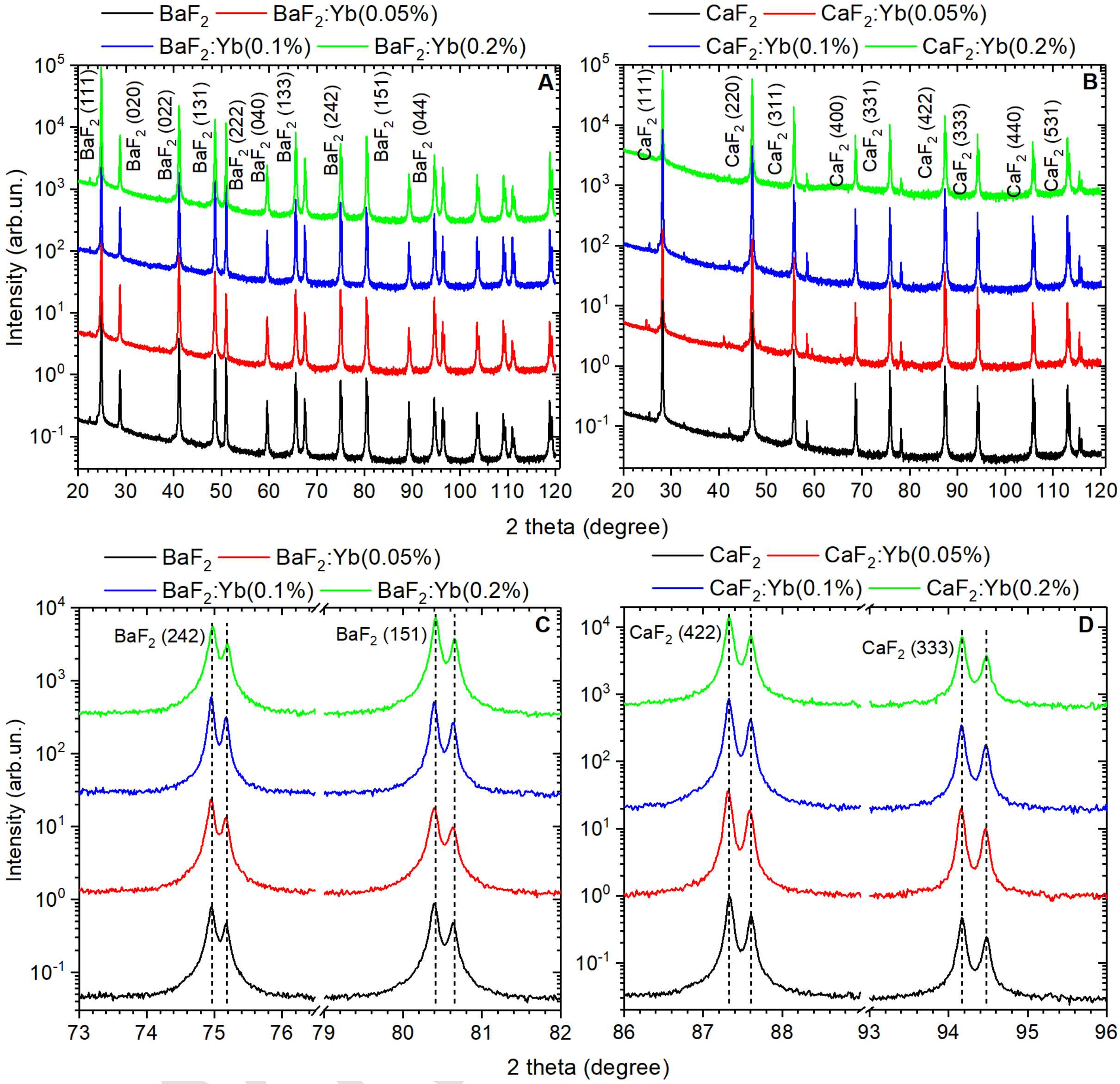

Fig. 2. A - XRD patterns of the $BaF_2$:Yb samples (see the legend). B - XRD patterns of the $CaF_2$:Yb samples (see the legend). C - chosen XRD peaks of the $BaF_2$:Yb samples (see the legend). D - chosen XRD peaks of the $CaF_2$:Yb samples (see the legend). Vertical dashed lines are eye-catching indicators demonstrating no shift of the peaks upon Yb doping. The most pronounced diffraction peaks are indexed with the corresponding Miller indices (hkl).

The lattice parameters were calculated and listed in Table 2.

Table 2. Lattice parameters and of the $BaF_2$:Yb and $CaF_2$:Yb samples.

| Sample | Lattice parameter, Å |
|---|---|
| $BaF_2$ | 6.202 ± 0.002 |
| 0.05 mol% $YbF_3$:$BaF_2$ | 6.202 ± 0.002 |
| 0.1 mol% $YbF_3$:$BaF_2$ | 6.202 ± 0.002 |
| 0.2 mol% $YbF_3$:$BaF_2$ | 6.200 ± 0.002 |
| $CaF_2$ | 5.465 ± 0.002 |
| 0.05 mol% $YbF_3$:$CaF_2$ | 5.466 ± 0.002 |
| 0.1 mol% $YbF_3$:$CaF_2$ | 5.465 ± 0.002 |
| 0.2 mol% $YbF_3$:$CaF_2$ | 5.465 ± 0.002 |

The lattice parameter remained practically constant upon the varying Yb content in both sets of samples within experimental uncertainty (±0.002 Å) (see Table 2). The near-constant lattice parameters across all doping levels indicate that Yb ions are substitutionally incorporated without causing measurable bulk lattice expansion or contraction in both types of fluorides studied. This stability supports the absence of large-scale structural phase changes (chosen peaks in Fig. 2C,D practically do not change their position upon Yb doping) or secondary phase formation, aligning with the transmittance data below and confirming high crystal quality.

In addition, Raman spectroscopy (the spectra are shown in Fig. 3) revealed sharp peaks at 248 cm$^{-1}$ and 324 cm$^{-1}$, corresponding to $BaF_2$ and $CaF_2$ lattice modes, respectively.

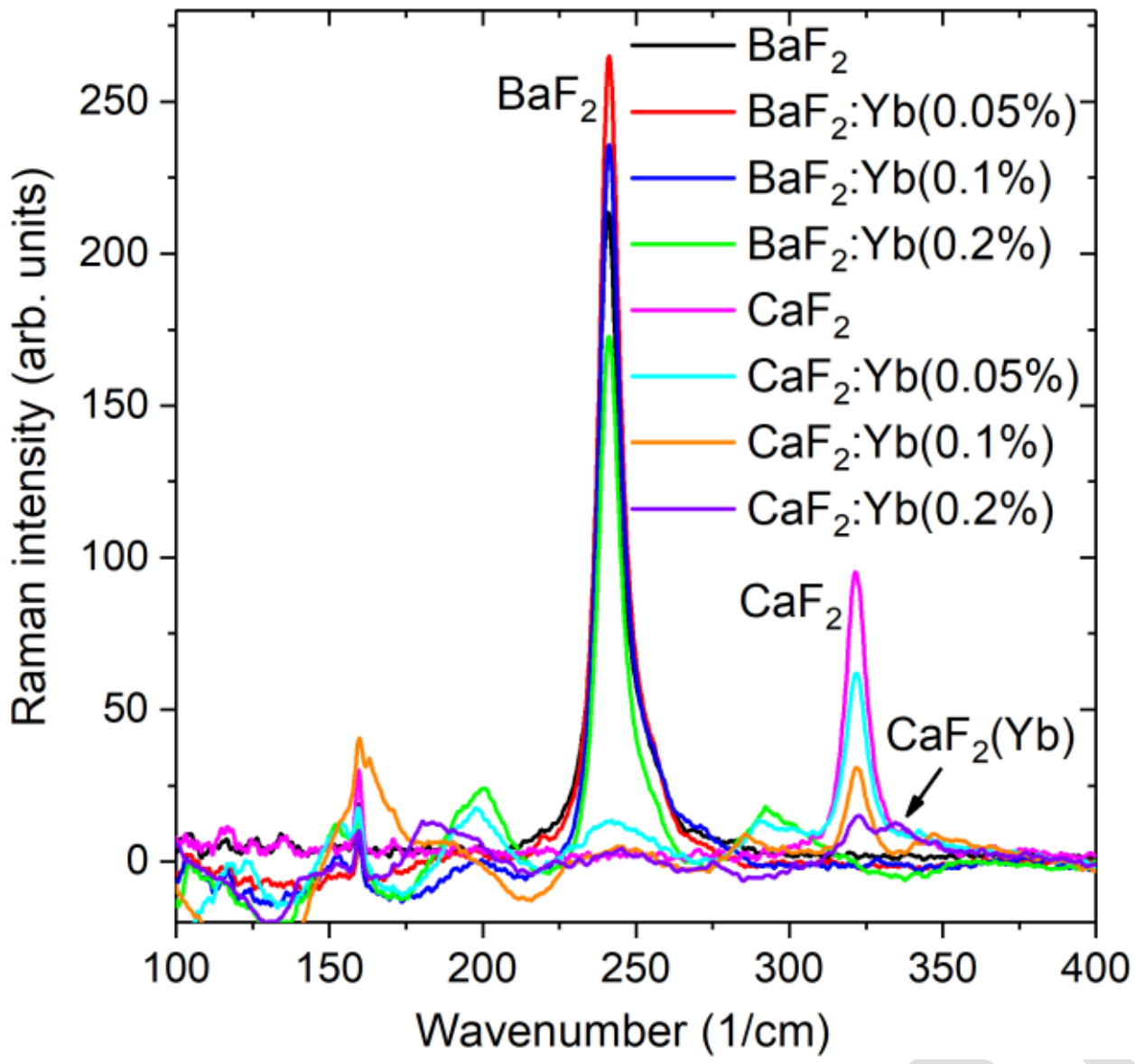

Fig. 3. Raman spectra of the Ca(Ba)$F_2$:Yb(0, 0.05, 0.1, 0.2%) samples. The peak at 248 cm$^{-1}$ belongs to the $BaF_2$ phase, whereas the peak at 324 cm$^{-1}$ originates from the $CaF_2$ phase. The peak at approximately 335 cm$^{-1}$ can be related to the $CaF_2$ sublattice perturbed by Yb, marked as $CaF_2$(Yb). The rest of the peaks may be attributed to $YbF_3$ or oxygen-halide-related phases of unknown origin. Negative peaks in the region 100-150 cm$^{-1}$ are due to the baseline subtraction.

The intensity of the peaks is changing upon Yb doping. The most pronounced changes were observed in the case of the $CaF_2$ - the larger the Yb content, the lower the peak. Moreover, the new peak at approximately 335 cm$^{-1}$ is observed at the 0.2% doping level. It can be related to the $CaF_2$ sublattice perturbed by Yb marked as $CaF_2$(Yb) as it is only 9 cm$^{-1}$ shifted with regard to the typical $CaF_2$ peak at 324 cm$^{-1}$. This suggests local symmetry lowering and the formation of Yb-induced local lattice perturbations in the $CaF_2$ host.

While XRD confirms long-range structural stability, Raman spectroscopy reveals local lattice perturbations, indicating that Yb incorporation induces short-range distortions without affecting the average crystal structure. There were also additional peaks that emerged exclusively in Yb-doped samples, suggesting the presence of trace $YbF_3$ and unidentified oxygen-halide phases. These may originate from Yb-rich clusters or surface reactions, as fluorides are known to undergo partial substitution of oxygen for fluorine upon environmental exposure. To check this XPS analysis of surface chemical composition was provided.

Negative-going features observed in some spectra arise from baseline subtraction and normalization procedures applied during data processing and do not represent physical negative intensities. These artifacts do not affect the determination of peak positions or the interpretation of relative spectral trends. The low-frequency Raman region (100-150 cm$^{-1}$) is particularly sensitive to baseline correction due to its proximity to the Rayleigh scattering line and the relatively low intensity of phonon modes. As a result, minor imperfections in background subtraction may lead to apparent distortions or negative-going features in the spectra.

While X-ray diffraction confirms that the average crystal structure remains unchanged within experimental uncertainty, Raman spectroscopy reveals local lattice distortions induced by Yb incorporation. The appearance of the mode near ~335 cm⁻¹ therefore reflects short-range perturbations of the $CaF_2$ sublattice rather than a global structural modification.

*3.2 Surface elemental composition*

High resolution (HR) XPS spectra were measured in the two sets of samples: undoped and Yb doped $BaF_2$ (Set 1) and $CaF_2$ (Set 2) single crystals. The comparison of normalized Ba 2p and Cl 2p spectra on example of the samples for Set 1 are shown in Fig. 4. According to [41], the Yb 4d peak should be located at 182.4 eV. Also, no clear signal of Yb 4f (3 eV) or Yb 4p ($4p_{1/2}$ 389 eV and $4p_{3/2}$ 341 eV) peaks were observed in survey XPS spectra. The energy loss peaks from Ba signal were observed in the region from 182 to 190 eV. In addition, the valence band region (from -5 to 23 eV) and region from ca. 340 - 395 eV in the Set 1 were scanned with HR resulting in the same outcome - no evidence for the presence of Yb 4f or Yb 4p peaks (spectra not shown here). As for the Set 2, XPS analysis also did not reveal presence of any Yb signal. The absence of detectable Yb signals is consistent with the low doping concentration (≤0.2 mol%), which is below the detection limit of XPS (see Experimental). At the same time the Ca 2p or Ba 2p peaks as well as F 1s peak were detected in the HR XPS spectra of both sets.

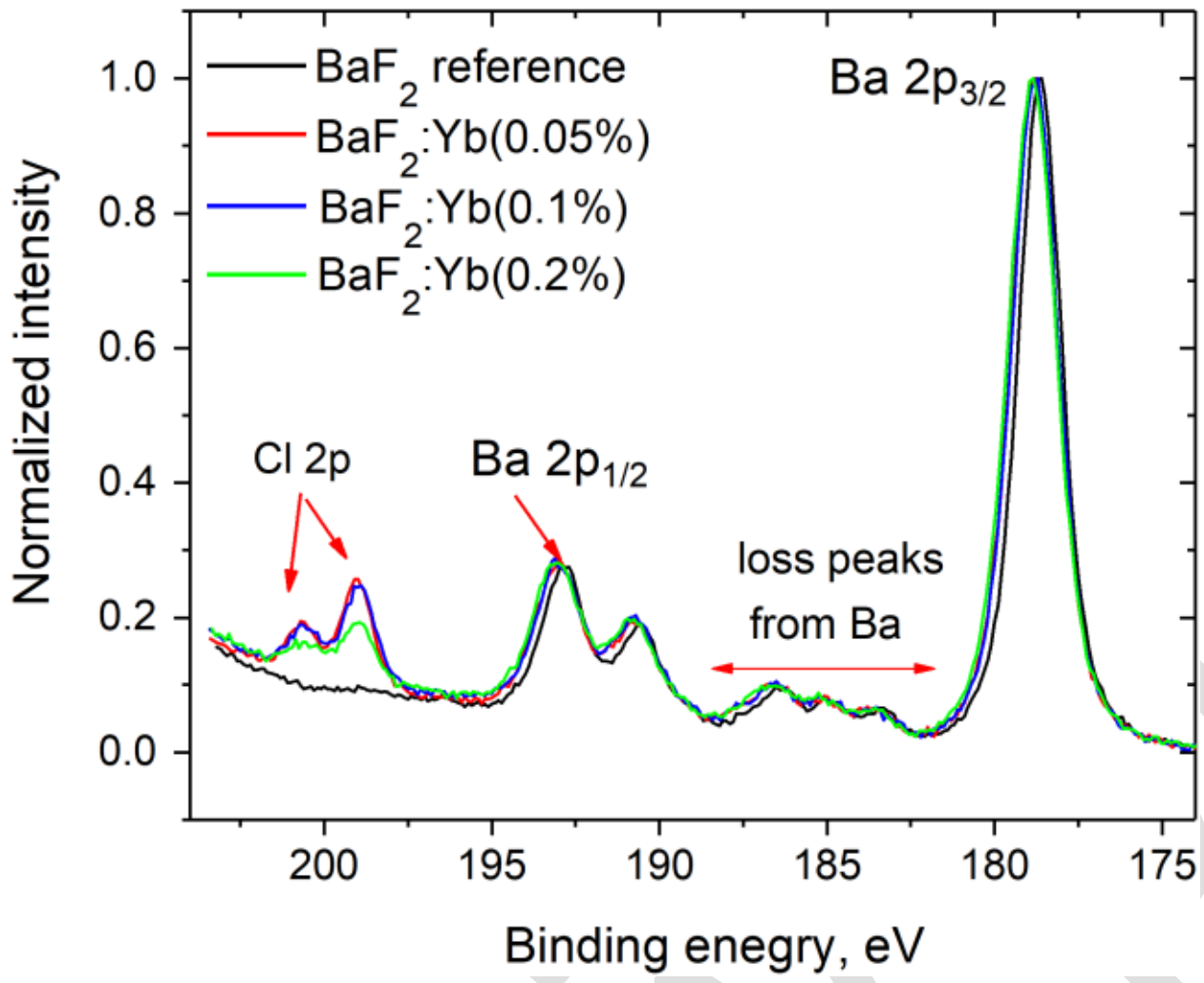

Fig. 4. The comparison of normalized Ba 2p and Cl 2p spectra of the samples for Set 1. No such signals were observed in the $CaF_2$ set of samples.

Moreover, a weak Ca signal, corresponding to a Ca concentration below 1 at.%, was detected in Set 1 samples, where no calcium was expected, suggesting possible contamination during sample preparation (remained in a crucible). Along with the Ca, Ba and F the Cl signal was detected only in the Yb doped $BaF_2$ samples. The atomic concentrations of chemical elements detected on the surface of the Set 1 and 2 were calculated from HR XPS spectra and listed in Table 3.

Table 3. The atomic concentration of chemical elements on the surface of samples calculated from HR XPS spectra

| Sample name | Atomic concentration of chemical elements (*n*), at.% | | | | | | | |
|---|---|---|---|---|---|---|---|---|
| | C | O | Ba | F | *n*(F)/*n*(Ba) | Ca | *n*(F)/*n*(Ca) | Cl |
| $BaF_2$ | 36.7 | 4.8 | 18.3 | 40.0 | 2.19 | 0.2 | - | - |
| $BaF_2$:Yb(0.05%) | 39.5 | 10.8 | 16.9 | 30.0 | 1.78 | 0.4 | - | 2.4 |
| $BaF_2$:Yb(0.1%) | 44.6 | 9.7 | 14.4 | 28.3 | 1.97 | 0.8 | - | 2.2 |
| $BaF_2$:Yb(0.2%) | 47.2 | 11.8 | 13.9 | 25.6 | 1.84 | 0.2 | - | 1.3 |
| $CaF_2$ | 28.4 | 3.9 | - | 42.5 | - | 25.2 | 1.67 | - |
| $CaF_2$:Yb(0.05%) | 26.0 | 3.6 | - | 44.0 | - | 26.4 | 1.67 | - |
| $CaF_2$:Yb(0.1%) | 22.9 | 3.1 | - | 46.4 | - | 27.6 | 1.68 | - |
| $CaF_2$:Yb(0.2%) | 22.1 | 3.1 | - | 46.9 | - | 27.9 | 1.68 | - |

All samples contained also relatively large amount of carbon and some amount of oxygen on the surface.

The origin of carbon probably originates from ambient environment where the carbon always exists in dust. Surprisingly, the surface carbon concentration in Set 1 varied with the increasing trend from 36.7 at.% in the undoped $BaF_2$ to 47.2 at.% in the $BaF_2$:Yb(0.2%). In contrast to the $BaF_2$, the carbon concentration had descending tendency upon the Yb content changing in Set 2 from 28.4 at.% in the undoped $CaF_2$ to 22.1 at.% in the $CaF_2$:Yb(0.2%). Moreover, in average, the carbon content was about twice as higher in the Set 1 as compared to Set 2. Similar trends were observed for the oxygen content indicating that the C-O bonds are created. At the same time, at least partial substitution of oxygen for fluorine at the surface cannot be excluded as demonstrated for $LiCaAlF_6$ [45]. The variation in carbon content is attributed to differences in surface reactivity and adsorption behavior between the two host materials leading to increasing upon the Yb doping level in the Set 1 and decreasing in the Set 2. This can be explained by the interstitial ion creating extra charge, especially at the surface, existence. The observed differences likely reflect different surface reactivity and adsorption behavior of the two fluoride hosts.

Barium and fluorine content is decreasing in the series of Yb doping levels 0-0.2 at.% from 18.3 to 13.9 at.% for Ba and 40 to 25.6 at.% for F. This can be explained by the Ca presence and the Yb substitution for Ba thus creating an extra positive charge ($\mathrm{Yb}_{\mathrm{Ba}}^{+}$) in the case of $Yb^{3+}$ and neutral ($\mathrm{Yb}_{\mathrm{Ba}}^{0}$) in the case of $Yb^{2+}$. To compensate for the mentioned extra positive charge barium vacancies ($V_{Ba}$) should be created or an interstitial fluorine $\mathrm{F}_{i}^{-}$ may appear (or both).

The decrement in case of barium content is in average about 1 at.% while in the case of fluorine it is about 5 at.% whereas the F to Ba content ratio is also strongly changing without any clear tendency (Table 3) in the series of Yb doping levels 0-0.2 at.%. The most drastic changes in the case of the F content (lowering from 40 to 30 at.%) appear when moving from 0 to 0.05% Yb doping level. All of these observations are consistent with the formation of fluorine vacancies ($V_F$) along with the $V_{Ba}$ existence influenced by the $Yb^{3+}$ content. Despite the Yb signals were below detection thresholds, the consistent trends in surface elemental ratios, especially the F/Ca and F/Ba ratios (see Table 3), offer indirect evidence of charge-compensating defect formation. These may involve interstitial fluorine, oxygen substitution, or vacancy complexes - defect motifs that have been previously predicted and experimentally validated in similar systems [45].

Calcium and fluorine content in Set 2 is increasing in the series of Yb doping levels 0-0.2 at.% from 25.2 to 27.9 at.% for Ca and 42.5 to 46.9 at.% for F. The increment in case of calcium and fluorine content is in average about 1 at.% and the F to Ca content ratio is practically constant value (Table 3) in the series of Yb doping levels of 0-0.2 at.%. To explain this in comparison with $BaF_2$ having opposite trends, the ionic radii of Ca, Ba and Yb should be considered. The ionic radii of $Ca^{2+}$, $Ba^{2+}$, $Yb^{2+}$ and $Yb^{3+}$ in the eightfold coordinated environment (considering the cubic crystal structure of the Ba(Ca)$F_2$ with Ca(Ba)$F_8$ cubes) is 1.120, 1.420, 1.140 and 0.985 Å, respectively [46]. The ionic radii of $Ca^{2+}$ and $Yb^{2+}$ are very similar with the difference of only 0.020 Å, thus leading to the negligibly small lattice distortion at the $Yb^{2+}$ substitution for $Ca^{2+}$. Obviously, the smaller $Yb^{3+}$ will result in the more pronounced impact on the lattice when embedded at $Ca^{2+}$ site as the ionic radii difference in this case is 0.135 Å (approximately 7 times ($0.135/0.02 \approx 7$) larger than that for the $Yb^{2+}$). Therefore, one may expect a higher probability of $Yb^{2+}$ stabilization in the $CaF_2$ host compared to $BaF_2$. These observations may indicate partial stabilization of $Yb^{2+}$ species in $CaF_2$ due to the close ionic radius match between $Ca^{2+}$ and $Yb^{2+}$.

The ionic radii of $Ba^{2+}$ and $Yb^{2+}$ are not as similar as in the case of $Ca^{2+}$ with the difference of 0.28 Å, thus leading to the relatively large lattice distortion even at the $Yb^{2+}$ substitution for $Ba^{2+}$. The difference in ionic radii in this case is about twice of that for the $Yb^{3+}$ and $Ca^{2+}$ in the $CaF_2$. The ionic radii difference in the case of $Ba^{2+}$ and $Yb^{3+}$ is even larger having the value of 0.57 Å, twice as larger as compared to the $Yb^{2+}$. Therefore, the domination of $Yb^{2+}$ over the $Yb^{3+}$ is expected in $BaF_2$ (Set 1) as well. However, it is not that pronounced as compared to $CaF_2$ (Set 2). Therefore, the higher $Yb^{3+}$ content in this case will lead to the larger alteration of the $BaF_2$ lattice as also described above.

Based on this analysis, one may expect a decrease of the Ca vacancies in $CaF_2$ (Set 2) as compared to $BaF_2$ (Set 1). Possibly, Yb also pushes Ca to the surface while $Yb^{3+}$ attracts more fluorine thus preserving constant F to Ca content ratio (see Table 3).

The evidence of oxygen and Cl impurities points to limited but detectable phase segregation at higher doping levels of Yb and correlates well with Raman measurements. The detected chlorine is most likely related to residual contamination from the graphite crucible, which had previously been used for the growth of chlorides. To clarify the situation with Yb incorporation into the $BaF_2$ and $CaF_2$ hosts, EPR measurements were made and described below.

### *3.3 Peculiarities of Yb incorporation*

EPR spectra measured in the as grown ytterbium doped $BaF_2$ at 20 K - the temperature optimal for the $Yb^{3+}$ signals observation, are shown in Fig. 5A.

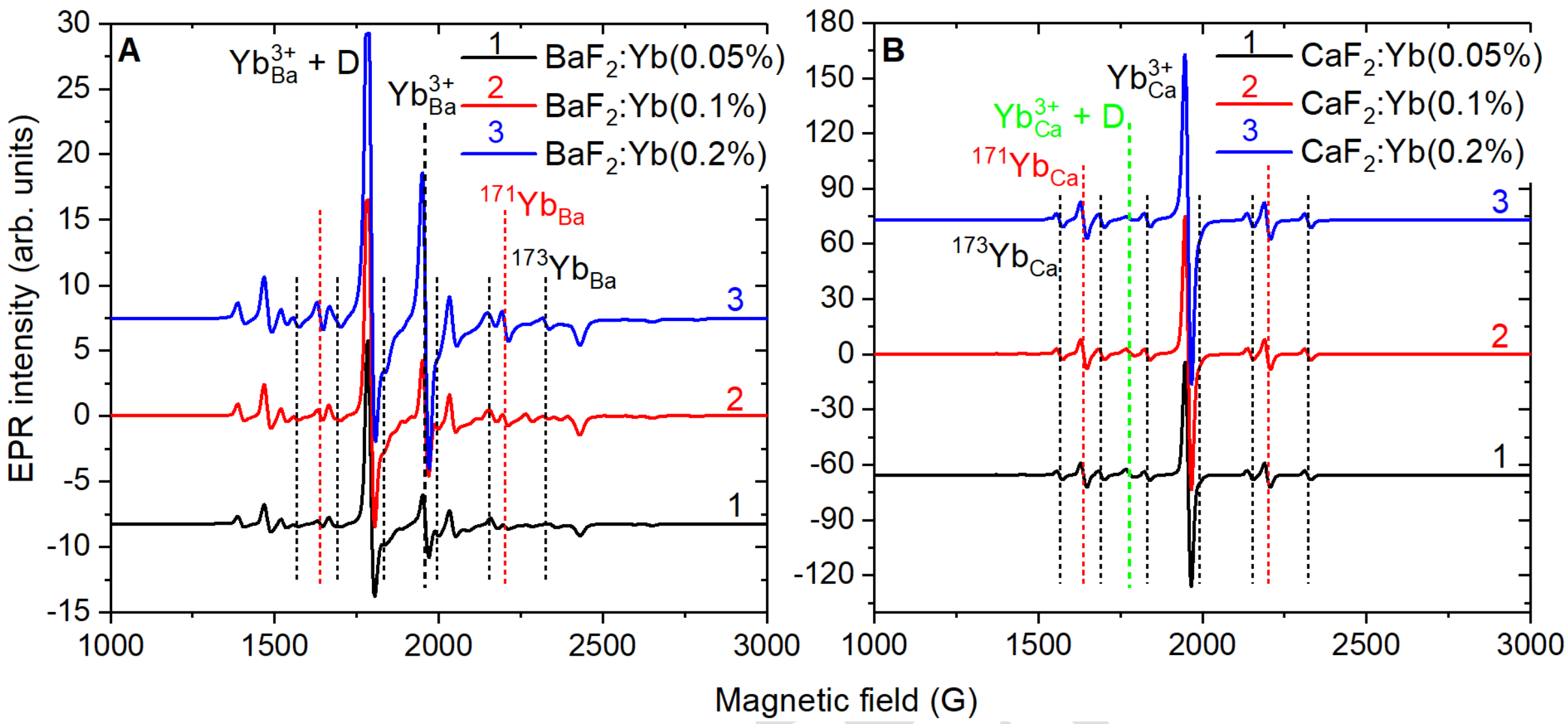

Fig. 5. A - EPR spectra measured in the $BaF_2$:Yb(0.05, 0.1, 0.2%) as also listed in the legend. $Yb^{3+}_{Ba}$ and $^{171,173}Yb_{Ba}$ indicate the signals originating from $Yb^{3+}$ occupying Ba site. The $^{171,173}Yb_{Ba}$, in particular, stand for the hyperfine interaction of the electron spin S = 1/2 of the single unpaired electron found in the $4f^{13}$ outer shell of the $Yb^{3+}$ and two isotopes $^{171}Yb$ and $^{173}Yb$ having natural abundance of ca. 14 and 16%, respectively. The $Yb^{3+}_{Ba} + D$ and the rest of the weaker resonances indicate the signals originating from $Yb^{3+}$ occupying Ba site perturbed by some imperfection nearby (D). The weaker resonances originate from the $^{171}Yb$ and $^{173}Yb$ as well. B - EPR spectra measured in the $CaF_2$:Yb(0.05, 0.1, 0.2%) as also listed in the legend. By analogy with the $Yb^{3+}_{Ba}$ in the panel A, the $Yb^{3+}_{Ca}$ and $^{171,173}Yb_{Ca}$ indicate the signals originating from $Yb^{3+}$ occupying Ca site and the corresponding hyperfine interaction with the $^{171,173}Yb$ isotopes. Again, the very weak $Yb^{3+}_{Ca} + D$ indicate the signals originating from $Yb^{3+}$ occupying Ca site perturbed by some imperfection nearby (D). The rest of the even much weaker resonances originating from the hyperfine interaction with the $^{171,173}Yb$ were not detected due to the low natural abundance of the isotopes. Numbers 1-3 indicate different spectra distinguished by color (also in the legends).

The signals were complex composed of the two strong resonance lines at the *g* factors 3.42 (~1960 G) and 3.75 (~1789 G) and about ten times weaker satellite lines symmetrically distributed around them. The line observed at the $g = 3.42$ and its satellites exhibited isotropic properties. The satellites formed one sextet and one doublet. Considering ytterbium doping and the temperature of observation mentioned above (there was no significant resonances above 20 K) one may conclude that these resonances originate from $Yb^{3+}$. Sextet is produced by the hyperfine interaction with $^{173}Yb$ nucleus (nuclear spin $I = 5/2$; natural abundance is about 16%) while the doublet is the result of the hyperfine interaction with $^{171}Yb$ (nuclear spin $I = 1/2$; natural abundance is about 14%) [47] (see Fig. 5A). The line observed at the $g = 3.75$ and its satellites exhibited strongly anisotropic properties typical for powder spectra [48,49]. These resonances had the same temperature dependence appearing in the similar spectral range as the $g = 3.42$ - centers resonances mentioned above. It is noteworthy, that there is only one cation site in the $BaF_2$ structure for the $Yb^{3+}$ to occupy. Therefore, the anisotropic signals were related to $Yb^{3+}$ located at the barium site perturbed by some imperfection nearby as compared to the origin of the isotropic signal. The isotropic and anisotropic signals were thus designated as $Yb^{3+}_{Ba}$ and $Yb^{3+}_{Ba} + D$. Integrated EPR intensities were evaluated to enable quantitative comparison of $Yb^{3+}$ concentration within each set of samples and between the two hosts. They are listed in Table 4.

Table 4. Double-integrated EPR signal intensities for Yb-doped $BaF_2$ and $CaF_2$ samples evaluated for $Yb^{3+}$ centers (Fig. 5).

| Host | Yb doping level | Center | Double integral |
|---|---|---|---|
| $BaF_2$ | 0.05% | $Yb^{3+}_{Ba} + D$ | $5.650 \times 10^7$ |
| $BaF_2$ | 0.1% | $Yb^{3+}_{Ba} + D$ | $8.530 \times 10^7$ |
| $BaF_2$ | 0.2% | $Yb^{3+}_{Ba} + D$ | $1.330 \times 10^8$ |
| $BaF_2$ | 0.05% | $Yb^{3+}_{Ba}$ | $7.210 \times 10^6$ |
| $BaF_2$ | 0.1% | $Yb^{3+}_{Ba}$ | $1.370 \times 10^7$ |
| $BaF_2$ | 0.2% | $Yb^{3+}_{Ba}$ | $3.930 \times 10^7$ |
| $CaF_2$ | 0.05% | $Yb^{3+}_{Ca} + D$ | $5.458 \times 10^6$ |
| $CaF_2$ | 0.1% | $Yb^{3+}_{Ca} + D$ | $4.739 \times 10^6$ |
| $CaF_2$ | 0.2% | $Yb^{3+}_{Ca} + D$ | $2.696 \times 10^6$ |
| $CaF_2$ | 0.05% | $Yb^{3+}_{Ca}$ | $1.890 \times 10^8$ |
| $CaF_2$ | 0.1% | $Yb^{3+}_{Ca}$ | $2.390 \times 10^8$ |
| $CaF_2$ | 0.2% | $Yb^{3+}_{Ca}$ | $3.010 \times 10^8$ |

For quantitative comparison, the EPR spectra were evaluated using double integration over selected resonance regions. The resonance lines centered at approximately ~1800 G and ~1961 G were attributed to perturbed and unperturbed $Yb^{3+}$ centers, respectively. The integration ranges were chosen to fully cover the corresponding resonance features while minimizing contributions from the baseline and adjacent spectral components. Specifically, the ~1800 G line was integrated over 1720-1870 G for $BaF_2$ samples and over a narrower range of 1750-1800 G for $CaF_2$ samples due to its reduced intensity and overlap with background features. The ~1961 G line was integrated over 1925-1995 G for all samples.

The hyperfine structure of $Yb^{3+}$ ions was not explicitly taken into account in the integration procedure in order to avoid additional complexity in the quantitative evaluation. The hyperfine interaction leads to splitting of the EPR lines for isotopes with non-zero nuclear spin ($^{171}Yb$, I = 1/2, 14.3% abundance; $^{173}Yb$, I = 5/2, 16.1% abundance), while the remaining isotopes are EPR-silent with respect to hyperfine splitting. As a result, the hyperfine contribution is distributed proportionally across all samples and does not affect the comparative analysis of double-integrated intensities.

The intensity ratios of the $Yb^{3+}_{Ba}$ to $Yb^{3+}_{Ba} + D$ as well as the $Yb^{3+}_{Ca}$ to $Yb^{3+}_{Ca} + D$ signals in all the samples studied (for the Yb doping levels 0.05, 0.1 and 0.2%) were calculated and listed in Table 5.

Table 5. Intensity ratio $I(Yb^{3+}_{Ba,Ca})/I(Yb^{3+}_{Ba,Ca} + D)$ derived from double-integrated EPR signals for Yb-doped $BaF_2$ and $CaF_2$ samples.

| Host | Yb doping level | $I(Yb^{3+}_{Ba,Ca})/I(Yb^{3+}_{Ba,Ca} + D)$ |
|---|---|---|
| $BaF_2$ | 0.05% | 0.128 |
| $BaF_2$ | 0.1% | 0.160 |
| $BaF_2$ | 0.2% | 0.296 |
| $CaF_2$ | 0.05% | 34.6 |
| $CaF_2$ | 0.1% | 50.4 |
| $CaF_2$ | 0.2% | 111.7 |

The pronounced increase of $I(Yb^{3+}_{Ca})/I(Yb^{3+}_{Ca} + D)$ in $CaF_2$ with increasing Yb concentration suggests progressive stabilization of regular $Yb^{3+}$ lattice sites relative to defect-perturbed centers. Moreover, these results suggest that increasing Yb concentration reduces the relative stability of the perturbed site ($Yb^{3+}_{Ba} + D$) compared to the unperturbed configuration.

To prove correctness of the EPR signal origin guess and to have better insight into the Yb incorporation, the corresponding $Yb^{3+}$ spectra were thoroughly numerically analysed for the case of the $BaF_2$:Yb(0.2%) sample (the strongest EPR signals were observed there). The experimental and calculated EPR spectra are shown in Fig. 6A.

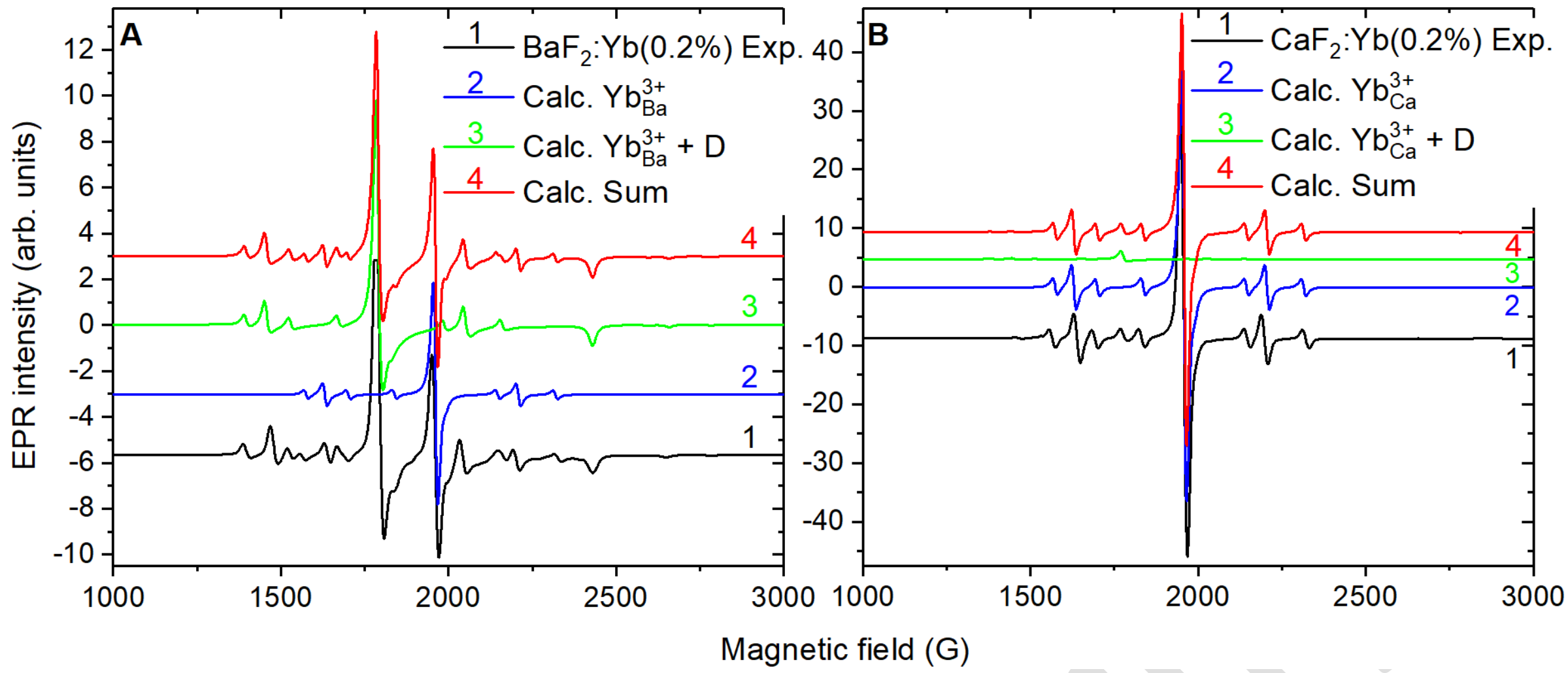

Fig. 6. A - EPR spectra measured (Exp.) and calculated (Calc.) for the $BaF_2$:Yb(0.2%). B - EPR spectra measured in the $CaF_2$:Yb(0.2%). The signals from $Yb^{3+}_{Ba(Ca)}$ and $Yb^{3+}_{Ba(Ca)}$ + D (D is some imperfection nearby) are also indicated in the legends. Numbers 1-4 indicate different spectra distinguished by color (also in the legends).

The experimental spectrum measured in the $BaF_2$:Yb(0.2%) has been approximated with the calculated one which is the combination of two signals. The following spin-Hamiltonian has been used to calculate both:

$$\hat{\mathbf{H}} = \beta\hat{\mathbf{S}}\hat{\mathbf{g}}\mathbf{B} + \hat{\mathbf{S}}\hat{\mathbf{A}}\hat{\mathbf{I}}, \quad (1)$$

where $\beta, \hat{\mathbf{S}}, \hat{\mathbf{I}}, \hat{\mathbf{g}}, \hat{\mathbf{A}}, \mathbf{B}$ are Bohr magneton, vector electron spin and nuclear spin operators, **g** tensor and tensor of hyperfine interaction, and magnetic field. The fit is excellent (see Fig. 6A). From this follows that the origins of the resonances were determined correctly. The fitting parameters determined (**g** and hyperfine tensor components for the $^{173}$Yb - the components for the $^{171}$Yb can be calculated multiplying those of the $^{173}$Yb by 3.81, the $^{171}$Yb to $^{173}$Yb Larmor frequencies ratio [40]) are listed in Table 6.

Table 6. Spin-Hamiltonian (Eq. 1) parameters. The **g** tensor components ($g_{1\text{-}3}$) determination error was ± 0.02. The hyperfine tensor components determination error was ± 20 MHz.

| Center | $g_1$ | $g_2$ | $g_3$ | $A_1$, MHz | $A_2$, MHz | $A_3$, MHz |
|---|---|---|---|---|---|---|
| $Yb^{3+}_{Ba}$ | 3.42 | 3.42 | 3.42 | 710 | 710 | 710 |
| $Yb^{3+}_{Ba}$ + D | 3.75 | 3.75 | 2.76 | 800 | 800 | 580 |
| $Yb^{3+}_{Ca}$ | 3.43 | 3.43 | 3.43 | 710 | 710 | 710 |
| $Yb^{3+}_{Ca}$ + D | 3.78 | 3.78 | 2.41 | 800 | 800 | 480 |

The values determined are typical for $Yb^{3+}$ in the case of both centers. The **g** and hyperfine tensors of the $Yb^{3+}_{Ba}$ + D are axial. Interestingly, the average value calculated from the principal **g** tensor values is $g_{av} = (g_1 + g_2 + g_3)/3 = 3.42$ which is equal to the g factor value of the $Yb^{3+}_{Ba}$. The average hyperfine tensor value $A_{av} = (A_1 + A_2 + A_3)/3 = 727$ MHz. Again, this is very close or even equal within the given error of determination (Table 6) to the hyperfine constant of the $Yb^{3+}_{Ba}$. All of these indicates electron density redistribution within the $Yb^{3+}$ ion upon the local symmetry lowering. Note, that the **g** and hyperfine tensor values are under the influence of local crystal field symmetry and strength (regulated by the bond length) [49,50]. Considering the cubic local symmetry of the Ba site ($BaF_8$) [51], the shift in the **g** tensor values of the $Yb^{3+}_{Ba}$ + D as compared to $Yb^{3+}_{Ba}$ can be explained by the changing the symmetry from cubic to tetragonal by elongation or squeezing of the $BaF_8$ cube, or off-centering of $Yb^{3+}$ along one of the cubic axes as the center symmetry should be axial due to the axial **g** and hyperfine tensors (Table 6). The most logical explanation to this phenomenon can be the existence of the interstitial fluorine anion next to the $\mathrm{Yb_{Ba}^{3+}}$ appearing on the axis undergoing changes resulting in the Yb-F bond length.

The situation is completely different in the $CaF_2$ isostructural with $BaF_2$. The corresponding EPR spectra are shown in Fig. 5B. Again, there were two different $Yb^{3+}$ related signals $\mathrm{Yb_{Ca}^{3+}}$ and $\mathrm{Yb_{Ca}^{3+}} + \mathrm{D}$. The former one originates from the unperturbed Ca site in the cubic $CaF_2$ structure. This was confirmed by the isotropic origin of this signal. The second signal was produced by the $Yb^{3+}$ but somehow perturbed as its spectral position is different as compared to the $\mathrm{Yb_{Ca}^{3+}}$ (Fig. 5A). Note, that, again, there is the only one cation site in the $CaF_2$ structure for the $Yb^{3+}$ to occupy. The double integrals as well as the intensity ratios of the $\mathrm{Yb_{Ca}^{3+}}$ to $\mathrm{Yb_{Ca}^{3+}} + \mathrm{D}$ signals in all the samples studied (for the Yb doping levels 0.05, 0.1 and 0.2%) are listed in Tables 4,5. However, the intensity of the $\mathrm{Yb_{Ba}^{3+}} + \mathrm{D}$ signal was dropping upon the Yb doping level (Table 4). All of these indicates that oppositely to the $BaF_2$ the $\mathrm{Yb_{Ca}^{3+}} + \mathrm{D}$ becomes less preferable under the increased Yb doping level as compared to the unperturbed site. Considering direct proportionality between EPR intensity and paramagnetic specie content [48,49] and based on Tables 4,5 one may conclude that there is much more of $Yb^{3+}$ incorporated into the $CaF_2$ host as compared to $BaF_2$, preferably, unperturbed.

To prove correctness of the EPR signal origin guess and to have better insight into the Yb incorporation in $CaF_2$ host, the corresponding $Yb^{3+}$ spectra were thoroughly numerically analysed for the case of the $CaF_2$:Yb(0.2%) sample (the strongest EPR signals were observed there and, moreover, the same doping level was considered in the case of $BaF_2$ samples). The experimental and calculated EPR spectra are shown in Fig. 6B.

The experimental spectrum measured in the $CaF_2$:Yb(0.2%) has been approximated with the calculated one which is the combination of two signals. The spin-Hamiltonian in Eq. 1 has been used to calculate both. The fit is excellent (see Fig. 6B). From this follows that the origins of the resonances were determined correctly. The fitting parameters determined (**g** and hyperfine tensor components) are listed in Table 6 as well.

The values determined are typical for $Yb^{3+}$ in the case of both centers. The **g** and hyperfine tensors of the $\mathrm{Yb_{Ca}^{3+}} + \mathrm{D}$ are axial. The average value calculated from the principal **g** tensor values is $g_{av} = (g_1 + g_2 + g_3)/3 = 3.32$ which is close to the g factor value of the $\mathrm{Yb_{Ca}^{3+}}$. The average hyperfine tensor value $A_{av} = (A_1 + A_2 + A_3)/3 = 693$ MHz. Again, this is very close or even equal within the given error of determination (Table 6) to the hyperfine constant of the $\mathrm{Yb_{Ca}^{3+}}$. All of these indicates electron density redistribution within the $Yb^{3+}$ ion upon the local symmetry lowering. Note, that the **g** and hyperfine tensor values are under the influence of local crystal field symmetry and strength (regulated by the bond length) [49,50]. Considering the cubic local symmetry of the Ca site ($CaF_8$) [52], the shift in the **g** tensor values of the $\mathrm{Yb_{Ca}^{3+}} + \mathrm{D}$ as compared to $\mathrm{Yb_{Ca}^{3+}}$ can be explained by the changing the symmetry from cubic to tetragonal by elongation or squeezing of the $CaF_8$ cube, or off-centering of $Yb^{3+}$ along one of the cubic axes as the center symmetry should be axial due to the axial **g** and hyperfine tensors (Table 6). The most logical explanation to this phenomenon can be the existence of the interstitial fluorine anion next to the $\mathrm{Yb_{Ca}^{3+}}$ appearing on the axis undergoing changes resulting in the Yb-F bond length. These are the same phenomena as described above for the $BaF_2$:Yb(0.2%). The difference in the spin-Hamiltonian parameters of Yb in $CaF_2$ is obviously very small as compared to the $BaF_2$ as these two crystals have the same symmetry so the crystal field strength can be expected to have small deviation moving from one crystal lattice to another.

Ytterbium also influences charge trapping processes occurring in the $BaF_2$ (all doping levels starting from 0 up to 0.2 %) whereas in the case of the $CaF_2$ this could only be observed for the 0.2% doping level. The signals created after the X-ray irradiation in the mentioned samples are shown in Fig. 7A.

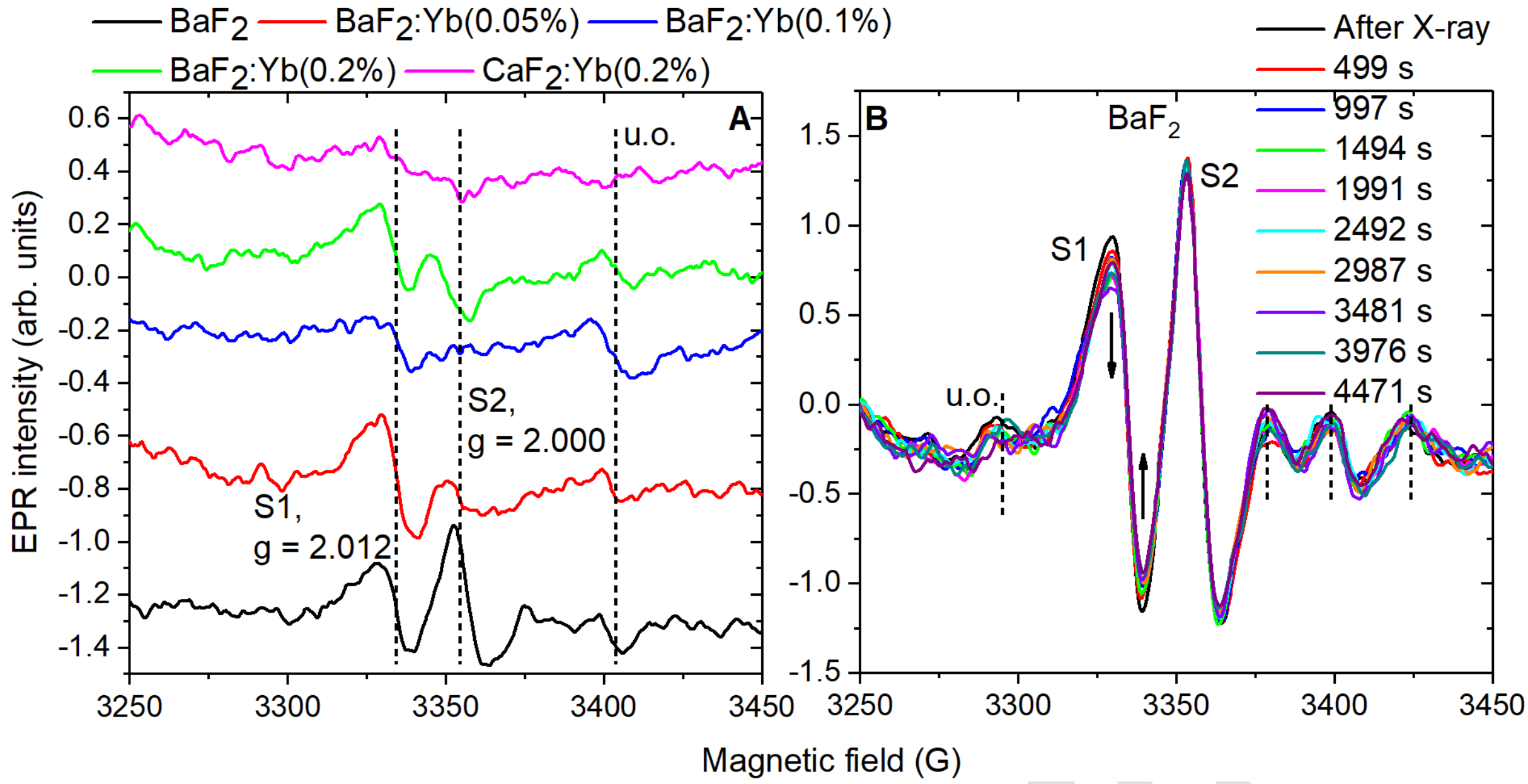

Fig. 7. A - EPR spectra measured in the $BaF_2$:Yb(0.05, 0.1, 0.2%) and $CaF_2$:Yb(0.2%) samples (as also listed in the legend) after the X-ray irradiation. S1,2 and the corresponding g factors indicate the two X-ray induced signals and their spectral positions. B - time dependence of the S1,2 signals after the X-ray irradiation measured in the $BaF_2$ sample. The time delays in the seconds range are listed in the legend. "u.o." indicates the signals of unknown origin present in the samples before the X-ray irradiation. It remains unchanged after the X-ray irradiation.

The X-ray induced signals were S1,2. They have different origin as the S1 to S2 intensity ratio is changing upon Yb doping level (2:3 (0%), 1:3 (0.05%), 1:1 (0.2%)). Note, that the S2 signal disappears in the $BaF_2$:Yb(0.1%). The S1 to S2 ratio cannot be determined in the case of the $CaF_2$:Yb(0.2%) as the signals are too weak and too broaden there (Fig. 7A). All of these, the g factors indicated in Fig. 7A and the absence of any other signals like the ones originated from the interaction of the electron spin and the nuclear spin of the intrinsic and/or nuclei found nearby for the S1,2 signals evidence for the signals origin rather to be oxygen-like centers ($O^-$ in the case of S1 and $O_2^-$ (pseudomolecular ion) in the case of the S2 signal) at the materials surface. To study their stability, the S1,2 signals time evolution was measured and shown in Fig. 7B. The S1 seems to have lower stability as compared to the S2 one - it decays with larger rate. However, this difference is very small (Fig. 7B).

The contrasting evolution of EPR signal intensities in $BaF_2$ and $CaF_2$ indicates differing defect dynamics. In $BaF_2$, the increase in perturbed site population with doping suggests enhanced local lattice distortion or clustering, while in $CaF_2$, the stability of unperturbed sites reflects a more favourable lattice environment for Yb incorporation. Considering room temperature charge trapping processes occurring almost exclusively in the $BaF_2$ host (Fig. 7A) one may conclude that namely the existence of $\mathrm{Yb}_{\mathrm{Ba}}^{3+} + \mathrm{D}$, especially, the stabilizing defect D is responsible for this. These findings corroborate the PL data showing broader crystal field effects in $CaF_2$ and support the conclusion that host lattice identity governs both electronic and optical defect interactions.

*3.4 Optical properties*

To complement the structural and compositional analyses, a series of optical measurements, including photothermal deflection spectroscopy, transmittance spectroscopy, and infrared photoluminescence were carried out to elucidate the influence of Yb incorporation on the optical properties of $BaF_2$ and $CaF_2$ host lattices.

### 3.4.1 Transmittance Spectroscopy

Transmittance spectra measured in both sets of $CaF_2$ and $BaF_2$ samples are shown in Fig. 8A. The $Yb^{3+}$ bands appearing at around 1.32 eV as the $^2F_{5/2} \rightarrow {}^2F_{7/2}$ transitions [12,53] as demonstrated in Fig. 8B (numbered 1-9). They are narrow and much weaker in the $BaF_2$ as compared to the strong and broadened transitions in the $CaF_2$ samples. This is in good correlation with EPR data above, indeed, the amount of $Yb^{3+}$ incorporated into the $CaF_2$ host is much larger than in the case of the $BaF_2$. In both $BaF_2$ and $CaF_2$ hosts, a slight but systematic decrease in transmittance was observed as the Yb content increased. Importantly, the absence of broad absorption bands or sharp dips due to scattering implies that the crystals retained their high optical quality and that Yb doping did not lead to significant phase segregation or defect-induced scattering centers. These findings corroborate the XRD results, which confirmed phase purity and indicated the preservation of crystalline homogeneity upon doping.

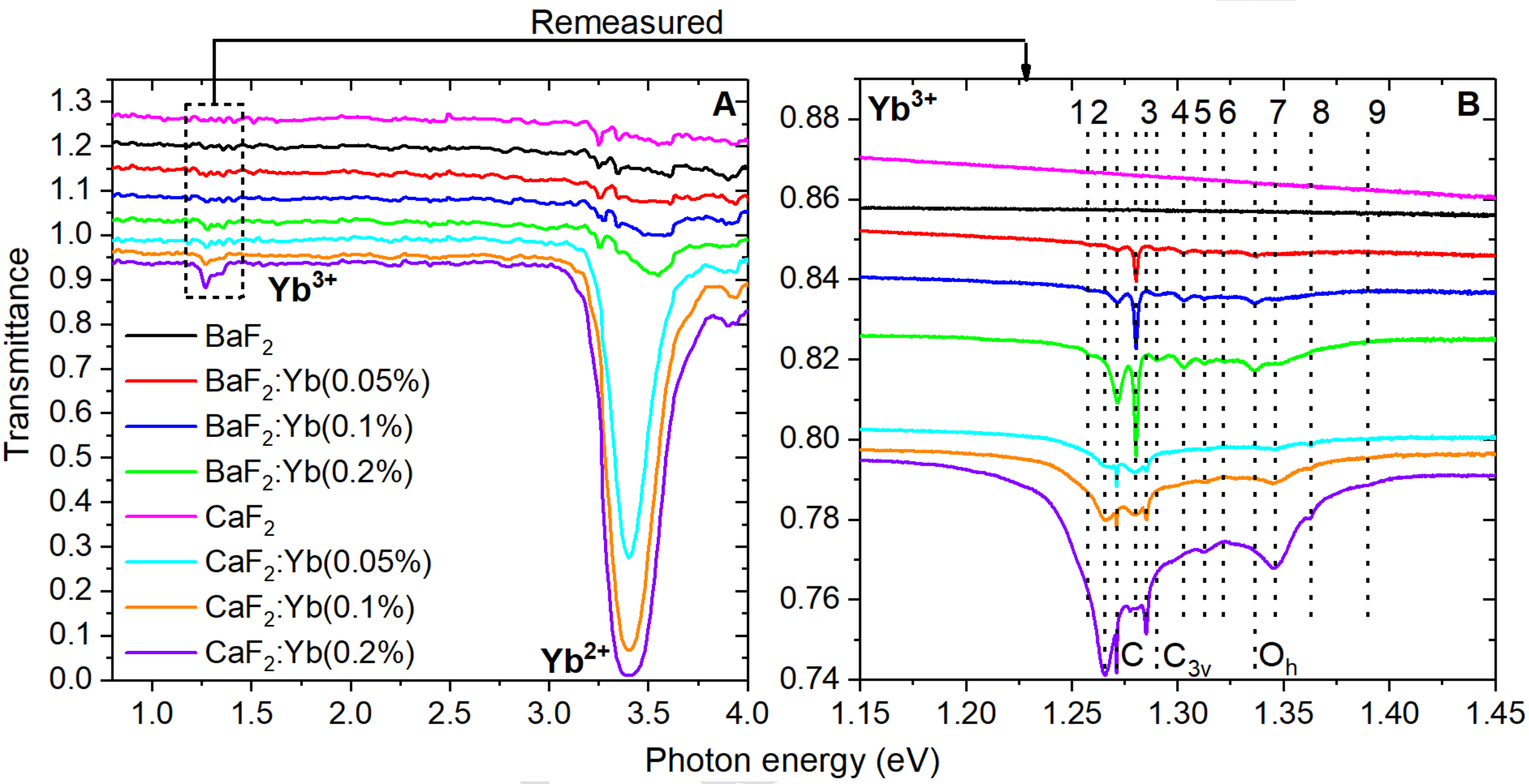

Fig. 8. A - Transmittance spectra of $BaF_2$:Yb and $CaF_2$:Yb samples (Yb contents are given in the legend). $Yb^{3+}$ as well as $Yb^{2+}$ transmittance bands are indicated. B - Close-up of the $Yb^{3+}$ transitions. Numbers 1-9 indicate specific transitions while C, $C_{3v}$, $O_h$ indicate transitions appearing due to the specific local symmetry of $Yb^{3+}$ created by the presence of interstitial fluorine [54].

The $Yb^{3+}$ transitions have complex and well-studied pattern. The presence of the bands indicated as $C_{3v}$ and $O_h$ (Fig. 8B) is ascribed to the presence of the interstitial fluorine for the charge compensation of $Yb^{3+}$. This is in good correlation with the EPR results discussed above. The suggestion of D being interstitial $F^-$ in the $Yb^{3+}_{Ba,Ca}$ + D center can be made.

The $Yb^{2+}$ transitions are typical for $Yb^{2+}$ as described in [12,55] in detail. They exhibit decrease upon the ytterbium doping level. Interestingly, the bands intensity is, in general, about one order of magnitude stronger in the $CaF_2$ as compared to the $BaF_2$ samples. Altogether with the trends for $Yb^{3+}$ discovered both in EPR and optical transmittance this leads to the conclusion that the segregation coefficient for the Yb doping must be much larger in the case of $BaF_2$ as compared to the $CaF_2$.

### 3.4.2 Peculiarities in the infrared photoluminescence

IR PL spectra were recorded under 810 nm excitation, a wavelength known to efficiently excite $Yb^{3+}$ ions via the $^2F_{7/2} \rightarrow {}^2F_{5/2}$ transition [12,53]. The spectra are shown in Fig. 9.

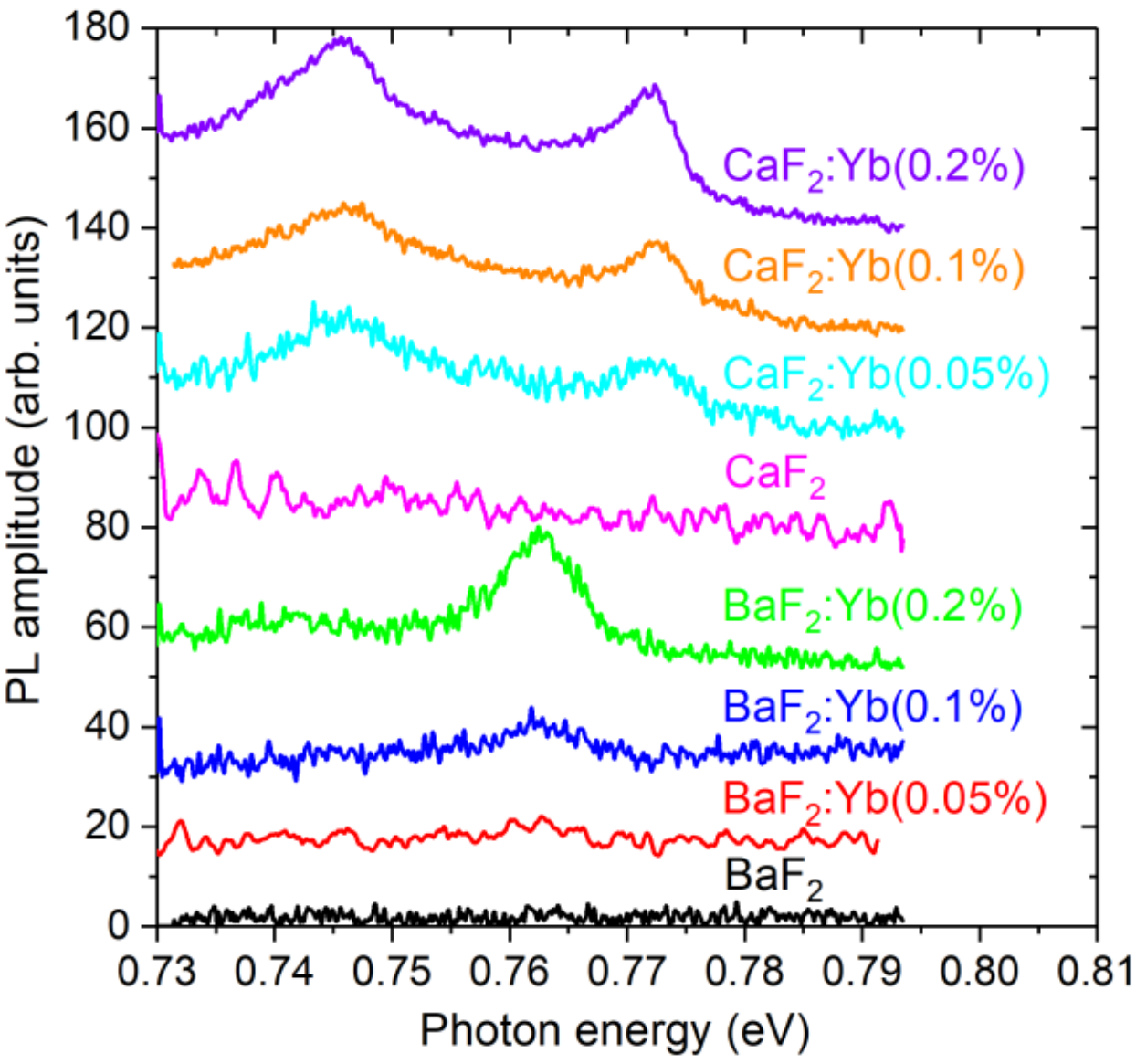

Fig. 9. Infrared photoluminescence spectra measured in the Ca(Ba)$F_2$:Yb(0, 0.05, 0.1, 0.2%) samples.

Distinct weak emission bands (never reported before) were detected exclusively in Yb-doped samples (their intensity was growing upon the Yb doping level), confirming the optical activity of the doped ions. In $BaF_2$:Yb, a single strong emission peak appeared at approximately 1628 nm (0.76 eV). In $CaF_2$:Yb, two emission bands were identified at 1608 nm (0.77 eV) and 1662 nm (0.75 eV), respectively. The splitting between the bands thus is 0.02 eV. Interestingly, the transition energy in the case of the $BaF_2$ is average of the transition energies of the peaks in the case of $CaF_2$. Moreover, it is approximately twice of the spin-orbit coupling constant of $Yb^{3+}$. Considering the spin-orbit coupling constant reaching in average $\lambda = 3000$ cm$^{-1}$ = 0.37 eV [56] and the energy splitting of $\Delta E = \lambda J = 0.37 \cdot 7/2 = 1.3$ eV, where $J$ is the quantum number corresponding to the total mechanical moment of the $Yb^{3+}$ ground state ($^2F_{7/2}$) [49], the observed emission bands can be assigned to the transitions within the $^2F_{5/2} \rightarrow {}^2F_{7/2}$ manifolds of $Yb^{3+}$. It should be noted that there are three levels in the case of the $^2F_{5/2}$ and four in the case of the $^2F_{7/2}$ so, in general, twelve transitions can be expected depending on the specific probabilities considering the symmetry of the local crystal field. The existence of one peak in the case of $BaF_2$ and two peaks in the case of $CaF_2$ can be explained based on the discussion above in the following way. The source of these peaks - $Yb^{3+}$ should be placed in the site with improved crystal field strength in the $CaF_2$ as compared to the $BaF_2$. Both $\mathrm{Yb}^{3+}_{\mathrm{Ba}}$ and $\mathrm{Yb}^{3+}_{\mathrm{Ca}}$ centers have the same cubic symmetry exhibiting practically no change in the $g$ factor the HF tensor (see Table 6). Therefore, they can be excluded from further consideration. The $\mathrm{Yb}^{3+}_{\mathrm{Ba,Ca}} + D$ have relatively large difference in the $g_3$ value (Table 6): $g_3(\mathrm{Yb}^{3+}_{\mathrm{Ba}} + D)$ - $g_3(\mathrm{Yb}^{3+}_{\mathrm{Ca}} + D) = 0.35$. The difference between the $g$ tensor values from Table 6 for the $\mathrm{Yb}^{3+}_{\mathrm{Ba,Ca}} + D$) is the following: $g_{1,2}(\mathrm{Yb}^{3+}_{\mathrm{Ba}} + D)$ - $g_3(\mathrm{Yb}^{3+}_{\mathrm{Ba}} + D) = 0.99$; $g_{1,2}(\mathrm{Yb}^{3+}_{\mathrm{Ca}} + D)$ - $g_3(\mathrm{Yb}^{3+}_{\mathrm{Ca}} + D)$ = 1.37. The ground state in both cases must be the superposition of several $|m_J\rangle$ $(-J \le m_J \le J)$ [49] ones: $\sum_{m_J} |m_J\rangle$. The increased difference between the $g_{1,2}$ and $g_3$ for the $\mathrm{Yb}^{3+}_{\mathrm{Ca}} + D$ as compare to the $\mathrm{Yb}^{3+}_{\mathrm{Ba}} + D$ confirms the removal of additional energy states $|m_J\rangle$ $(-J \le m_J \le J)$ by the splitting in the crystal field interrupted by the interstitial $F^-$. Moreover, the HF tensor components ($A_{1\text{-}3}$ in Table 6) differ by about 100 MHz for the $\mathrm{Yb}^{3+}_{\mathrm{Ba}} + D$ and $\mathrm{Yb}^{3+}_{\mathrm{Ca}} + D$ centers (they are large in the case of the $\mathrm{Yb}^{3+}_{\mathrm{Ca}} + D$ center) contributing to the better localization of the electron on the $Yb^{3+}$ ion to have better overlap between the electron and nuclear wavefunctions. All of these indicates the change in the local crystal field strength in the $\mathrm{Yb}^{3+}_{\mathrm{Ca}} + D$ as compared to the $\mathrm{Yb}^{3+}_{\mathrm{Ba}} + D$. This allows for the assumption that these two centers ($\mathrm{Yb}^{3+}_{\mathrm{Ba,Ca}} + D$) produce the observed IR PL bands, and the crystal field is indeed stronger in the case of the $\mathrm{Yb}^{3+}_{\mathrm{Ca}} + D$ thus resulting in the pronounced splitting of one band into two. Based on these considerations, schematically, the energy scheme is shown in Fig. 10.

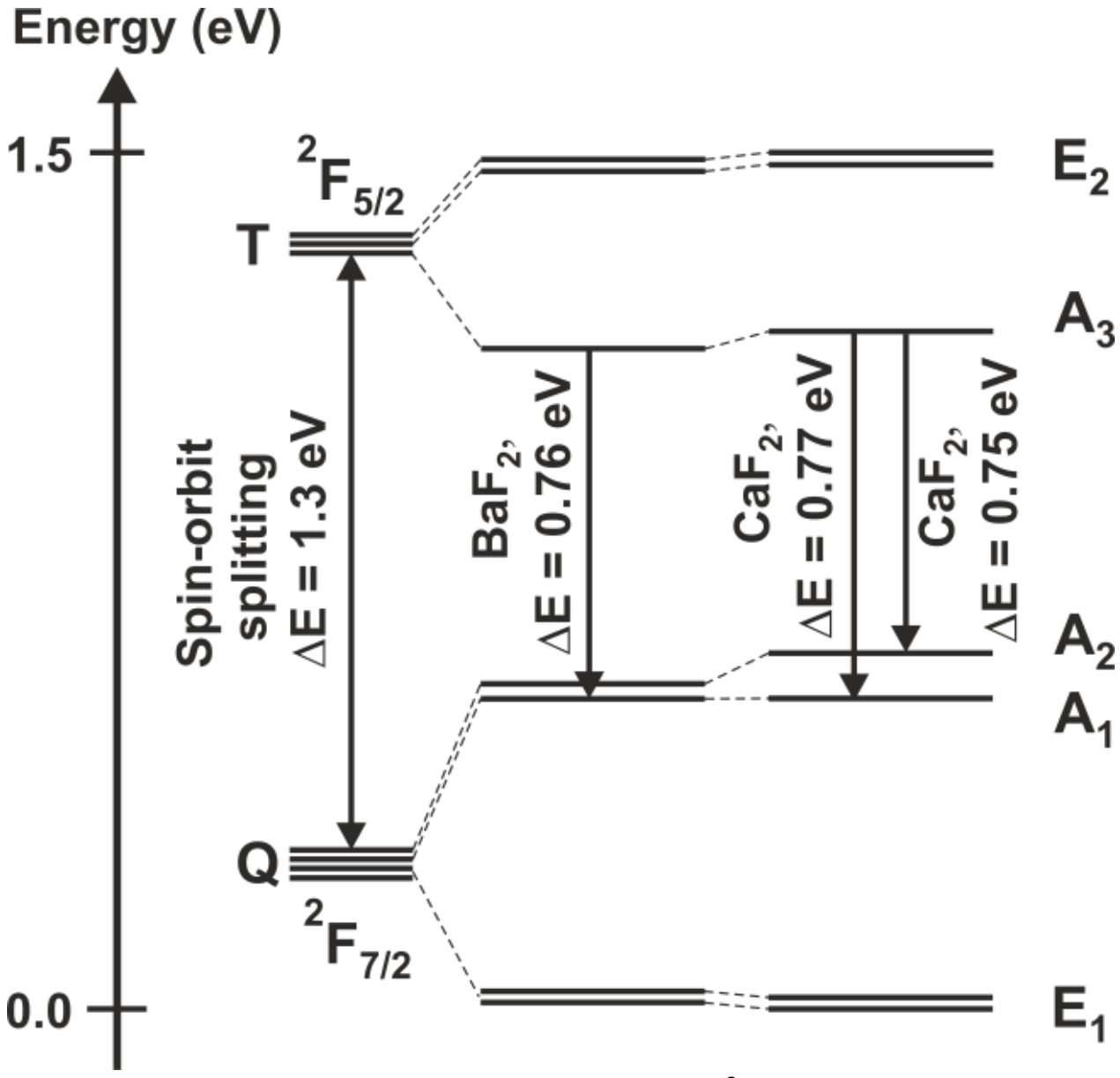

Fig. 10. Energy scheme for the $Yb^{3+}$ IR PL in Fig. 9B for $CaF_2$ and $BaF_2$.

It should be noted that the emission around ~980 nm represents the characteristic $Yb^{3+}$ transition ($^2F_{5/2} \rightarrow {}^2F_{7/2}$) and is commonly observed in a wide range of Yb-doped materials [57,58]. In contrast, the emission detected in the ~1620 nm region is not typical for $Yb^{3+}$ systems and is strongly influenced by the local environment and defect structure. To the best of our knowledge, such emission has not been previously reported for Yb-doped $BaF_2$ and $CaF_2$, making it a distinctive feature of the present study. In contrast to the conventional ~980 nm emission, which is largely insensitive to local structural variations, the ~1630 nm band appears to be strongly influenced by the local crystal field and defect environment. This makes it a sensitive probe of host-dependent effects and defect-mediated interactions.

Moreover, the observed near-infrared emission at ~1628 nm (~0.76 eV) falls within the technologically important spectral region around 1.6 µm. This wavelength range is particularly relevant for optical telecommunications, as it overlaps with the L-band (1565-1625 nm) and its extension, where silica optical fibers exhibit minimal transmission losses. Materials emitting in this region are therefore attractive for optical amplifiers, light sources, and integrated photonic devices operating in advanced communication systems.

The emission band is further structured due to local crystal field effects, exhibiting splitting into components at ~1608 nm (~0.77 eV) and ~1662 nm (~0.75 eV). This behavior can be attributed to Stark splitting of the $Yb^{3+}$ energy levels depending on the local symmetry of the lattice. Such spectral structuring may be advantageous for applications requiring broader or spectrally tunable emission, including wavelength-selective photonic systems and optical sensing.

In addition to telecommunications, emission in the ~1.6 µm range is relevant for infrared sensing and spectroscopy, where various molecular absorption features are present. This spectral region also lies within the eye-safe window, making it suitable for applications such as Light Detection And Ranging (LIDAR), range finding, and free-space optical communication. Furthermore, longer-wavelength NIR emission enables deeper penetration into biological tissues with reduced scattering, which is beneficial for biomedical imaging and diagnostic techniques.

Importantly, $CaF_2$ and $BaF_2$ single crystals provide a suitable platform for such applications due to their low phonon energy, high infrared transparency, and excellent chemical, thermal, and radiation stability. These properties ensure stable, reproducible, and efficient NIR emission, making these materials promising candidates for practical photonic applications in the ~1.6 µm spectral region.

**Conclusions**

This study presents a comprehensive analysis of Yb incorporation in $BaF_2$ and $CaF_2$ single crystals at low doping concentrations (0.05-0.2 mol%). Structural, electronic, and optical characterization techniques revealed distinct host-dependent behaviors in dopant charge state, site occupation, and defect interaction. While both hosts maintained cubic phase purity, EPR spectra indicated a higher fraction of perturbed $Yb^{3+}$ sites in $BaF_2$ with increasing concentration, suggesting local symmetry breaking. In contrast, $CaF_2$ exhibited more stable unperturbed sites, likely due to the close ionic radius match with $Yb^{2+}$ and a reduced need for charge compensation. XPS results confirmed surface defect evolution, including oxygen and chlorine presence, with host-specific trends in fluorine and carbon content. PDS and IR PL spectra provided clear evidence of optically active $Yb^{3++}$ centers, with $CaF_2$ showing crystal-field-induced splitting of emission peaks. Additionally, Raman spectra confirmed the emergence of minor $YbF_3$- and oxygen-halide-related vibrational modes in doped samples, indicating limited secondary phase formation and fluoride sensitivity to ambient oxygen. These observations are supported by XPS and EPR results and complement the host-specific emission behavior identified via IR PL. Together, these findings emphasize the need to consider both lattice geometry and environmental effects when designing Yb-doped fluoride systems for high-performance optical applications. The retention of high optical quality and luminescence efficiency in both hosts underscores their potential for quantum and photonic applications. Overall, $CaF_2$ emerges as a more favorable host for stabilizing Yb ions with minimal lattice distortion, while $BaF_2$ allows a broader spectrum of defect-driven site environments.

Across all characterization techniques, consistent trends emerge regarding host-dependent Yb behavior. XRD analysis confirms the structural phase stability, while XPS reveals the host-specific surface chemical composition and potential defect compensation mechanisms. EPR identifies local symmetry breaking and electronic defect states, particularly under X-ray excitation. These results converge with optical data, where $CaF_2$ offers a more uniform Yb environment and $BaF_2$ demonstrates broader site variability. This synergy across methods reinforces the conclusion that careful host selection is essential for tuning Yb charge states and defect interactions.

These findings lay the groundwork for tailoring rare-earth-doped fluoride crystals by host selection, enabling optimization of luminescent properties for lasers, scintillators, and quantum devices.

**Acknowledgment**
This work made use of the large research infrastructure CzechNanoLab (LM2023051).